\documentclass{jpp}
\usepackage{graphicx}

\usepackage{textgreek}

\usepackage[utf8]{inputenc}
\usepackage[T1]{fontenc}
\usepackage{amsmath}
\usepackage{hyperref}
\usepackage{array}
\usepackage{booktabs}
\usepackage{mathtools}

\newcolumntype{C}[1]{>{\centering\arraybackslash}p{#1}}

\hypersetup{
    colorlinks=true,
    linkcolor=blue, 
    citecolor=blue,
}
\usepackage{xcolor}
\let\oldS\S
\renewcommand{\S}{\textcolor{blue}{\oldS}}

\shorttitle{The Geometry of Flux Surfaces with Quasi-Poloidal Symmetry}
\shortauthor{R. Madan et al.}

\title{The Geometry of Flux Surfaces with Quasi-Poloidal Symmetry}

\author{Rishin Madan,\aff{1,2}
  \corresp{\email{rishin.madan@princeton.edu}}
  Wrick Sengupta,\aff{1} Elizabeth J. Paul,\aff{3} \newline Mohammed Haque,\aff{3}
  Richard Nies,\aff{4,5} José Luis Velasco,\aff{6} \and Amitava Bhattacharjee\aff{1}}

\affiliation{\aff{1}Department of Astrophysical Sciences, Princeton University,
Princeton, NJ 08543, USA
\aff{2}Princeton Plasma Physics Laboratory, Princeton, NJ 08540, USA
\aff{3}Columbia University, New York, NY 10027, USA
\aff{4}Rudolf Peierls Centre for Theoretical Physics, Parks Road, Oxford, OX1 3PU, UK
\aff{5}New College, Holywell Street, Oxford, OX1 3BN, UK
\aff{6}Laboratorio Nacional de Fusión, CIEMAT, 28040 Madrid, Spain}

\begin{document}

\maketitle

\begin{abstract}
Quasi-poloidal (QP) magnetic fields have desirable properties for confining plasma: no radial drift of guiding centres (with positive implications for neoclassical transport), zero Pfirsch-Schlüter current, and a lower level of damping for poloidal flows. Despite their attractive properties, QP fields are not amenable to the near-axis expansion, a major theoretical tool for understanding toroidal fields. In this paper, we provide a novel framework for defining and understanding QP flux surfaces. This framework relies on a simplification that transforms the task of finding a quasi-poloidal flux surface from a 3D problem to a 2D problem. This simplification also applies to asymmetric magnetic mirrors with desirable properties. We sketch how this 2D problem can form the basis of an efficient optimisation problem for finding QP flux surfaces. We leverage this 2D problem for theoretical understanding: for instance, we identify a route to finding QP flux surfaces that are naturally flat mirrors \citep{velasco2023robust_flat_mirror}. The reduced model is qualitatively checked against numerically optimised QP equilibria. These numerical solutions only satisfy QP approximately, but we predictably find that local discrepancies with the reduced model correspond to significant local QP errors, anomalous parallel currents, and field lines deviating from geodesics.

\end{abstract}
\section{Introduction}\label{sec: introduction}
The stellarator is a device that confines plasma by generating a toroidal magnetic field, envisaged for the purpose of energy production by controlled fusion  \citep{Spitzer58}. The shaping of the magnetic field critically affects properties of the plasma, such as confinement time and pressure profiles.

We focus on quasi-poloidally symmetric (QP) magnetic fields with negligible enclosed toroidal current. Such fields are a subset of quasisymmetric fields \citep{NUHRENBERG1988113, Boozer_1995,Rodriguez2020}, which have a symmetry of the magnetic field strength such that the guiding-centre motion of the particles possesses a conserved canonical momentum, analogous to the axisymmetric, tokamak case. Quasisymmetry itself is a special case of \textit{omnigeneity}, for which the bounce-averaged radial drift of the trapped particles vanishes \citep{hall1975}. When omnigenous fields have poloidally closed contours of the magnetic field strength, they are called quasi-isodynamic (QI) \citep{gori-lotz-nuehrenberg}. Quasi-poloidal symmetry is the particular case of quasisymmetry where the magnetic field strength is constant on poloidal loops, as opposed to toroidal (quasi-axisymmetry) or helical loops (quasi-helical symmetry). QP is thus also a special case of QI.

QP magnetic fields have advantages beyond those of other quasisymmetric and QI configurations. Given that the enclosed toroidal current of a QP flux surface vanishes, which is a good approximation for QP stellarators due to vanishing bootstrap and Pfirsch-Schlüter currents, the field lines are geodesics of the surface. Consequently, the radial drift of guiding centres vanishes pointwise \citep{palumbo1968_Closed_MHD}---the ideal scenario for collisionless particle confinement. Hence, there is no radial neoclassical transport to first order in guiding-centre motion. For the same reasons, the bootstrap current vanishes \citep{Landreman2012}. The Pfirsch-Schlüter current also vanishes (see, e.g., \S\ref{subsec: what is special}), resulting in zero \cite{shafranov_equilibrium_1963} shift and potentially higher beta limits. Reducing geodesic curvature has also been shown to improve electrostatic turbulence in QI stellarators \citep{Roberg-Clark_Xanthopoulos_Plunk_Stroteich_2026}.

Furthermore, QP has the potential to reduce anomalous transport. Uniquely, there is a lower level of damping for poloidal flows \citep{Shaing83}. Thus, QP allows for shearing flows that can possibly suppress turbulence \citep{Wobig_1999,terry00,Spong_2005,alcuson16}, though this is yet to be verified. As a final advantage of QP, there is evidence that second stability is possible in high-beta, compact, QP fields \citep{ware2004}, though this may not be a general feature of QP fields.

There has been recent success in optimising for QI fields with suitable properties for plasma confinement \citep{Goodman23,Sánchez_2023}. Thus, stellarators with QI fields are now tenable candidates for fusion reactors \citep{lion25,Hegna_2025}. In designing a reactor, it is useful to understand the landscape of viable magnetic fields. Understanding QP further progresses this goal, despite QP being a more restrictive assumption than QI.

Despite their attractive properties, quasi-poloidal magnetic fields are overlooked, often on the basis of two theoretical arguments. Firstly, there is a prevailing misconception (see, e.g., \citealt{garren_existence_1991}) that if the radial drift vanishes pointwise, then the magnetic field strength must be constant on the flux surface, which is unreasonably constraining. This is not true, however: QP flux surfaces with zero net toroidal current also have vanishing radial drift. As \cite{palumbo1968_Closed_MHD} first showed, given vanishing radial drift, the orthogonals to field lines must be symmetry directions of the field strength. If these orthogonals do not close on themselves, it does indeed follow that the field strength is a flux function. QP circumvents this argument, since the orthogonals to the field lines do indeed close on themselves. Granted, the only explicitly known class of magnetic fields that has geodesic field lines is the \cite{palumbo1968_Closed_MHD} magnetic fields, which do have constant field strength on the flux surfaces and are the only axisymmetric fields with this property \citep{SCHIEF_2003}. Nonetheless, the lack (or impossibility) of exact analytical QP solutions does not preclude the attainment of very accurate, numerically generated QP magnetic fields.

Secondly, exact QP is impossible close to the magnetic axis. The proof of this is reproduced in appendix \ref{apdx: impossibility of QP near axis}. However, this formal proof has limited relevance for practical stellarators. For low aspect ratio stellarators, the `near-axis' region is generally much smaller than the volume enclosed by the last closed flux surface, though the region of validity of near-axis can vary by configuration. For large aspect ratio stellarators, the level of QP-breaking around the axis can be made asymptotically small. Nonetheless, the proof indicates that we cannot replicate the success of directly constructing approximate fields using a near-axis expansion, as is possible for QI and other forms of quasisymmetry \citep{Plunk_Landreman_Helander_2019,Landreman_Sengupta_2019,Jorge_Plunk_Drevlak_Landreman_Lobsien_Camacho_Mata_Helander_2022,Camacho_Mata_Plunk_Jorge_2022,Rodríguez_2023,Goodman23}. In this work, we provide an alternative analytical tool for constructing and understanding QP flux surfaces.

The principal result of this paper is a reduction of the problem of finding QP surfaces. Generally, to find a flux surface with desired properties, one has to solve a 3D problem, as properties such as field strength and currents depend on field lines adjacent to the flux surface. In contrast, we show that one can solve a simpler 2D problem in order to produce individual QP flux surfaces. This simpler problem is realised through the \textit{surface equations}, presented in \S\ref{subsec: the surface equations}. This framework achieves many of the simplifying features of a near-axis expansion. Firstly, it provides a reduced parametrisation of a QP surface, since only three 2D functions satisfying the surface equations need to be specified. These three functions are curvature, torsion, and the `orthogonal distance' between field lines. Furthermore, it represents a simplification of the equilibrium problem, since only three equations need to be solved. This streamlines the direct approach, where one instead needs to specify three 3D functions, e.g. $\boldsymbol{r}(l,\alpha,\psi)$ for position vector $\boldsymbol{r}$ and coordinates $(l,\alpha,\psi)$, that solve an overdetermined set of equilibrium equations. In exchange for simplicity, this approach has limitations: it only guarantees a single QP surface and a surface found through this method need not enclose nested QP flux surfaces. 

The outline of this paper is as follows. In \S\ref{sec: moving frame representation}, we describe the magnetic field within a moving frame formalism. This is necessary for revealing the ensuing simplification. Readers primarily interested in results may skip the mathematical details of this section, but should note that much of the notation used in this paper is defined there. In \S\ref{sec: reduced model}, we first define a local generalisation to QP surfaces, termed \textit{local QP} (LQP) surfaces --- the exact relationship to QP is discussed in \S\ref{subsec: relation to QP}. In \S\ref{subsec: the surface equations}, the main result is presented: a reduced, decoupled model for LQP surfaces, termed the surface equations. In \S\ref{subsec: properties of surface equations}, we sketch how the surface equations can be used to formulate a simplified surface optimisation problem for finding QP surfaces and in \S\ref{subsec: what is special}, we discuss the physics of what makes QP flux surfaces susceptible to this simplification. In \S\ref{sec: possibility of irrational QP surfaces}, we note two canonical classes of solutions: generalised helicoids, with surfaces of revolution (magnetic mirrors) as a notable subtype, and Hasimoto surfaces, which are surfaces traced out by the self-induced motion of a vortex filament. In the case of toroidal Hasimoto surfaces, we develop a theory of the magnetic field strength (\S\ref{subsec: field strength on Hasimoto}): we show that such surfaces in vacuum are \textit{flat mirrors} \citep{velasco2023robust_flat_mirror} and derive governing equations for the field strength. In \S\ref{sec: numerical verification}, the surface equations are compared with numerical magnetic field configurations optimised for QP symmetry. We note that it is the surface equations which are directly checked, but not their application as a dimensional reduction. Local deviations from the surface equations correspond to where there is significant QP error (\S\ref{subsec: appearance of cusps}), anomalous parallel currents (\S\ref{subsec: non-zero parallel current}), and deviations of field lines from geodesics (\S\ref{subsec: When are Field Lines Sufficiently Geodesic}). Moreover, through the surface equations, we explain the prevalence of sharp cusps in these configurations (\S\ref{subsec: appearance of cusps}) and provide a possible explanation for high mirror ratios and narrow pinch points (\S\ref{subsec: high mirror ratios and narrow pinch points}) in certain types of optimized QP stellarators.

\section{A moving frame representation of a magnetic field}\label{sec: moving frame representation}

In this section, we consider a particular description of a solenoidal field $\boldsymbol{B}$. We work in a field-line-following coordinate system $(l,\alpha,\psi)$, where 
\begin{subequations}
    \begin{align}
   \frac{ \partial \boldsymbol{r}}{\partial l} &= \boldsymbol{\hat{t}}\coloneq\frac{\boldsymbol{B}}{B} ,\label{eq: covariant basis vector 1}\\
    \boldsymbol{B}&=\bnabla \psi \times \bnabla \alpha,
    \end{align}
\end{subequations}
where we define the position vector $\boldsymbol{r}$ and the tangential unit vector $\boldsymbol{\hat{t}}$. Such a field-line-following coordinate system always exists locally due to the \cite{Clebsch+1859+1+10} representation. In \S\ref{sec: reduced model}, we adopt a specific choice of field-line-following coordinates when we consider LQP surfaces.

\begin{figure}
    \centering
    \includegraphics[width=\linewidth]{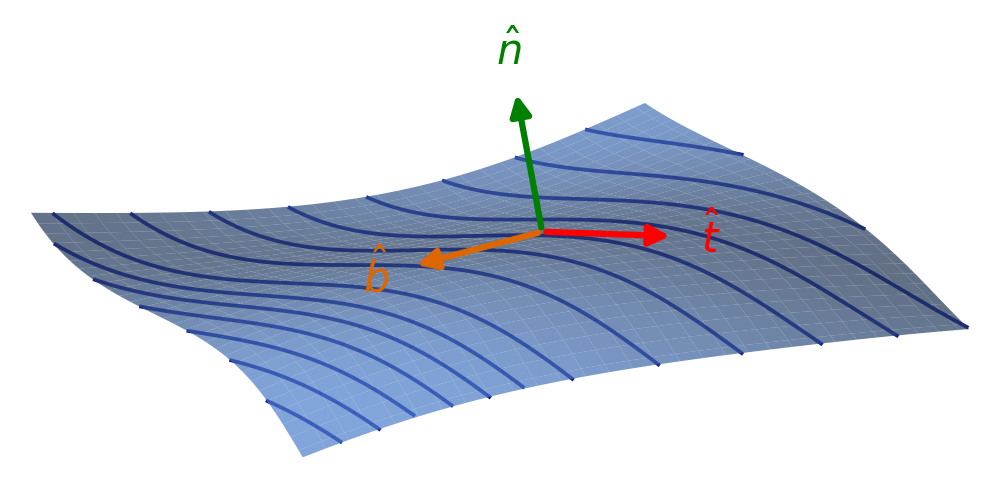}
    \caption{Illustration of the Darboux frame on a flux surface (constant $\psi$). $\boldsymbol{\hat{t}}$ is parallel to field lines (shown in dark blue) and lies in the tangent plane to the surface, $\boldsymbol{\hat{n}}$ is normal to the surface, and $\boldsymbol{\hat{b}}=\boldsymbol{\hat{t}}\times\boldsymbol{\hat{n}}$ also lies in the tangent plane to the surface.}
    \label{fig:darboux frame}
\end{figure}

We expand vectors in the orthonormal frame $(\boldsymbol{\hat{t}},\boldsymbol{\hat{n}},\boldsymbol{\hat{b}})$, where $\boldsymbol{\hat{n}}=\bnabla \psi / |\bnabla \psi |$ is the surface normal and the binormal is $\boldsymbol{\hat{b}} = \boldsymbol{\hat{t}} \times \boldsymbol{\hat{n}}$. This is the Darboux frame \citep{darboux2000lecons} of the surfaces given by the level sets of $\psi$, i.e., the flux surfaces, with the frame orientated along field lines. This is shown in figure \ref{fig:darboux frame}. Though we use the usual nomenclature, such flux surfaces are considered only locally at this stage.

We expand the covariant basis vectors in this frame:
\begin{subequations}\label{eqs: covariant basis vectors}
    \begin{align}
        \frac{\partial \boldsymbol{r}}{\partial \alpha} &= \eta \boldsymbol{\hat{t}} + \rho \boldsymbol{\hat{b}},\label{eqs: covariant basis vectors a}\\
        \frac{\partial \boldsymbol{r}}{\partial \psi} &= \lambda \boldsymbol{\hat{t}} + \mu \boldsymbol{\hat{n}}+\nu \boldsymbol{\hat{b}},\label{eqs: covariant basis vectors b}
    \end{align}
\end{subequations}
for some $\eta,\rho,\lambda,\mu$, and $\nu$. We have used the fact that $\partial_\alpha\boldsymbol{r}\cdot \bnabla \psi =0$. In terms of these variables, $B=1/\mu \rho$ and $\bnabla \psi = \boldsymbol{\hat{n}}/\mu$. 

The variables $\eta,\rho,\lambda,\mu$, and $\nu$ are not independent of each other: they constitute the metric in this coordinate system, so they are constrained by the fact that 3D Euclidean space is flat. Namely, the Riemann curvature tensor, which is a function of the metric, should vanish. Calculating such flatness conditions directly is algebraically cumbersome. Instead, we use the \textit{method of moving frames} \citep{darboux2000lecons,finikov_riemannian_2001}. Although seemingly circuitous, this method naturally provides the requisite variables for understanding the geometry of the flux surface as a surface embedded in 3D Euclidean space. Moreover, this method is utilised in \S\ref{sec: reduced model} for revealing a simplification for QP surfaces. In appendix \ref{apdx: riemann curvature}, we show the equivalence between the constraints that are derived below and the Riemann flatness conditions.

The Darboux frame $(\boldsymbol{\hat{t}},\boldsymbol{\hat{n}},\boldsymbol{\hat{b}})$ is orthonormal, smoothly varying in space, and is a complete basis, so we can define $\boldsymbol{K}=(K_1,K_2,K_3), \, \boldsymbol{\chi}=(\chi_1,\chi_2,\chi_3),$ and $\boldsymbol{\sigma}=(\sigma_1,\sigma_2,\sigma_3)$ such that 
\begin{subequations}\label{eqs: defs for K, chi, sigma}
    \begin{align}
        \frac{\partial}{\partial l}\begin{pmatrix}
            \boldsymbol{\hat{t}} \\
            \boldsymbol{\hat{n}} \\
            \boldsymbol{\hat{b}} \\
        \end{pmatrix}
        &= 
        \begin{pmatrix}
            0 & -K_3 & K_2 \\
            K_3 & 0 & -K_1 \\
            -K_2 & K_1 & 0
        \end{pmatrix}
        \begin{pmatrix}
            \boldsymbol{\hat{t}} \\
            \boldsymbol{\hat{n}} \\
            \boldsymbol{\hat{b}} \\
        \end{pmatrix}, \label{eqs: defs for K, chi, sigma a} \\
        \frac{\partial}{\partial \alpha}\begin{pmatrix}
            \boldsymbol{\hat{t}} \\
            \boldsymbol{\hat{n}} \\
            \boldsymbol{\hat{b}} \\
        \end{pmatrix}
        &= 
        \begin{pmatrix}
            0 & -\chi_3 & \chi_2 \\
            \chi_3 & 0 & -\chi_1 \\
            -\chi_2 & \chi_1 & 0
        \end{pmatrix}
        \begin{pmatrix}
            \boldsymbol{\hat{t}} \\
            \boldsymbol{\hat{n}} \\
            \boldsymbol{\hat{b}} \\
        \end{pmatrix}, \label{eqs: defs for K, chi, sigma b}\\
        \frac{\partial}{\partial \psi}\begin{pmatrix}
            \boldsymbol{\hat{t}} \\
            \boldsymbol{\hat{n}} \\
            \boldsymbol{\hat{b}} \\
        \end{pmatrix}
        &= 
        \begin{pmatrix}
            0 & -\sigma_3 & \sigma_2 \\
            \sigma_3 & 0 & -\sigma_1 \\
            -\sigma_2 & \sigma_1 & 0
        \end{pmatrix}
        \begin{pmatrix}
            \boldsymbol{\hat{t}} \\
            \boldsymbol{\hat{n}} \\
            \boldsymbol{\hat{b}} \\
        \end{pmatrix}.
    \end{align}
\end{subequations}
These equations can be thought of as a generalisation of the Frenet-Serret formulas \citep{Serret1851,Frenet1852}. Relative to the field lines on the flux surface, $-K_3$ is the normal curvature, $K_2$ is the geodesic curvature, and $K_1$ is the geodesic torsion \citep{eisenhart2013treatise}. The anti-symmetry of the matrices follows from the orthonormality of the frame.

Compatibility of these equations in flat space requires that the partial derivatives applied to the Darboux frame commute. For example, $\partial_l \partial_\alpha (\boldsymbol{\hat{t}},\boldsymbol{\hat{n}},\boldsymbol{\hat{b}}) = \partial_\alpha \partial_l (\boldsymbol{\hat{t}},\boldsymbol{\hat{n}},\boldsymbol{\hat{b}})$. This leads to
\begin{subequations}\label{eq: frame compatibility}
    \begin{align}
        \boldsymbol{\chi}_l - \boldsymbol{K}_\alpha &= \boldsymbol{K} \times \boldsymbol{\chi}, \label{eq: frame compatibility a}\\
        \boldsymbol{K}_\psi - \boldsymbol{\sigma}_l &= \boldsymbol{\sigma} \times \boldsymbol{K}, \label{eq: frame compatibility b} \\
        \boldsymbol{\sigma}_\alpha - \boldsymbol{\chi}_\psi &= \boldsymbol{\chi} \times \boldsymbol{\sigma}, \label{eq: frame compatibility c}
    \end{align}
\end{subequations}
where the subscripts denote partial derivatives. This is consistent with intuition in flat space: vectors parallel transported in closed loops should return to the same vector, so partial derivatives applied to vectors should commute.

Furthermore, $\boldsymbol{r}_{l\alpha}=\boldsymbol{r}_{\alpha l}$ gives
\begin{subequations}\label{eqs: lb compatibility}
    \begin{align}
        \eta_l &= K_2 \rho,\label{eq: lb compatibility a}\\
        \chi_3 &= K_3 \eta - K_1 \rho,\label{eq: lb compatibility b}\\
        \chi_2 &= K_2 \rho + \rho_l.\label{eq: lb compatibility c}
    \end{align}
\end{subequations}
Similarly, $\boldsymbol{r}_{l\psi}=\boldsymbol{r}_{\psi l}$ gives
\begin{subequations}\label{eqs: lpsi compatibility}
    \begin{align}
        \lambda_l &= K_2 \nu - K_3 \mu,\label{eq: lpsi compatibility a}\\
        \mu_l &= K_3 \lambda - K_1 \nu  -\sigma_3,\label{eq: lpsi compatibility b}\\
        \nu_l &= K_1\mu - K_2 \lambda +\sigma_2.\label{eq: lpsi compatibility c}
    \end{align}
\end{subequations}
Finally, $\boldsymbol{r}_{\alpha\psi}=\boldsymbol{r}_{\psi \alpha}$ gives
\begin{subequations}\label{eqs: bpsi compatibility}
    \begin{align}
        \eta_\psi - \lambda_\alpha &= \sigma_2 \rho + \chi_3 \mu - \chi_2 \nu,\label{eqs: bpsi compatibility a}\\
        \mu_\alpha &= \sigma_1 \rho - \sigma_3 \eta + \chi_3 \lambda - \chi_1 \nu, \label{eqs: bpsi compatibility b}\\
        \rho_\psi - \nu_\alpha &= \chi_2 \lambda - \chi_1 \mu -\sigma_2 \eta.\label{eqs: bpsi compatibility c}
    \end{align}
\end{subequations}

Finding $\eta,\rho,\lambda,\mu,\nu,\boldsymbol{K},\boldsymbol{\chi},$ and $\boldsymbol{\sigma}$ that satisfy equations \eqref{eq: frame compatibility}, \eqref{eqs: lb compatibility}, \eqref{eqs: lpsi compatibility}, and \eqref{eqs: bpsi compatibility} in some domain $\mathcal{D}_{(l,\alpha,\psi)}$ is equivalent to specifying a solenoidal field $\boldsymbol{B}$ in the real-space domain $\mathcal{D}_{\boldsymbol{r}}$ that corresponds to integrating the covariant basis vectors (equations \eqref{eq: covariant basis vector 1} and \eqref{eqs: covariant basis vectors}) over $\mathcal{D}_{(l,\alpha,\psi)}$. In the subsequent sections, we consider further properties of the magnetic field in terms of these variables. 

For future reference, we also write the current $\boldsymbol{J}=\bnabla \times \boldsymbol{B}$ as
\begin{multline}\label{eq: current}
    \boldsymbol{J} = \frac{1}{\mu \rho} \bigg[ \left( \left(\frac{1}{\mu \rho}\right)_\alpha -\left(\frac{\eta}{\mu \rho}\right)_l \right) \frac{\partial \boldsymbol{r}}{\partial \psi} \\+ \left(  \left(\frac{\eta}{\mu \rho}\right)_\psi -\left(\frac{\lambda}{\mu \rho}\right)_\alpha \right) \frac{\partial \boldsymbol{r}}{\partial l} + \left( \left(\frac{\lambda}{\mu \rho}\right)_l-\left(\frac{1}{\mu \rho}\right)_\psi   \right) \frac{\partial \boldsymbol{r}}{\partial \alpha} \bigg],
\end{multline}
which can be derived using the expression for the curl in general coordinates (see, e.g., Chapter 5 in \citealt{imbert-gerard_introduction_2024}).

\section{A reduced model for local QP surfaces}\label{sec: reduced model}

We define a \textit{local QP} (LQP) surface as a flux surface where
\begin{enumerate}
    \item the field lines are geodesics,
    \item the current normal to the surface is zero, and
    \item the current parallel to the field lines is zero.
\end{enumerate}These properties depend only on the local (albeit 3D) details of the magnetic field, whereas QP is a global property of a toroidal magnetic field. Nonetheless, LQP surfaces are the appropriate local generalisation of QP surfaces: force-balanced QP magnetic fields with zero enclosed toroidal current have LQP flux surfaces, with the precise relationship discussed in \S\ref{subsec: relation to QP}. We note that flux surfaces in magnetic mirror geometry can also be LQP surfaces.

Independently of the relationship with QP surfaces, LQP surfaces are worthy of consideration on physical grounds. That the current normal to the surface vanishes is a trivial consequence of magnetohydrostatic (MHS) force balance,
\begin{equation}\label{eq: MHS force balance}
    \boldsymbol{J}\times\boldsymbol{B}=\bnabla p(\psi),
\end{equation}
where we assume a scalar pressure. If the field lines are geodesics of the flux surface, the radial drift in the guiding centre approximation vanishes pointwise, resulting in no radial neoclassical transport and no bootstrap current to first order in guiding-centre motion. The Pfirsch-Schlüter current also vanishes for geodesic field lines, so the plasma does not generate its own parallel currents; hence, it is reasonable to further assume that the parallel current is zero. Indeed, this has its own merits: negligible parallel current improves MHD stability and is one of the intrinsic advantages of a stellarator over a tokamak \citep{Helander_2012}.

The main result of this section---and indeed of the paper---is introduced in \S\ref{subsec: the surface equations}, where we present a surprisingly simple model for producing LQP surfaces. We also sketch how this model can be used to independently generate QP surfaces (\S\ref{subsec: properties of surface equations}) and discuss the physics that makes LQP surfaces amenable to such simplification (\S\ref{subsec: what is special}).

\subsection{Relation to quasi-poloidal symmetry}\label{subsec: relation to QP}
We start by examining LQP surfaces. By definition, field lines are geodesics of the surface if the geodesic curvature $K_2$ vanishes. Physically, this is because the `acceleration' term $\partial \boldsymbol{\hat{t}}/\partial l$ has no component in the tangent plane of the surface, spanned by $(\boldsymbol{\hat{t}},\boldsymbol{\hat{b}})$, as can be seen from equation \eqref{eqs: defs for K, chi, sigma a}.  By equation \eqref{eq: lb compatibility a}, this implies $\eta_l=0$, and we can therefore change coordinates from $(l,\alpha)$ to $(l',\alpha)$ by shifting the origin of the field lines,
\begin{equation}\label{eq: l transformation}
    l' \coloneq l - \int^\alpha d\alpha' \, \eta(\alpha',\psi).
\end{equation}
It is noted that in a toroidal geometry, one has to exercise care in extending the $(l,\alpha)$ coordinates globally, but this transformation introduces no novel issues, since it is merely a statement of where we choose to define $l=0$ \citep{imbert-gerard_introduction_2024}. For instance, if $2\pi\psi$ is the enclosed toroidal flux, then we can restrict $\alpha \in [0,2\pi)$, and pick the lower limit of the integral in equation \eqref{eq: l transformation} to be $0$. Equation \eqref{eq: l transformation} then implies $\eta'=0$, where $\eta'$ is defined as in equation \eqref{eqs: covariant basis vectors} but now with respect to $(l',\alpha,\psi)$ coordinates. This choice of coordinates is a particular example of \textit{geodesic coordinates} (see, e.g., \citealt{eisenhart2013treatise}). The coordinates $l'$ and $\alpha$ are orthogonal on the surface, since $(\partial_\alpha\boldsymbol{r})_{l'} = \rho' \boldsymbol{\hat{b}}$ is perpendicular to $(\partial_{l'}\boldsymbol{r})_\alpha= \boldsymbol{\hat{t}}$. Henceforth, we work in such coordinates, so we drop the prime on $l',\eta'$, etc. Thus, the geodesic condition on LQP surfaces simply becomes $\eta=0$.

Additionally, by definition, the current normal to the LQP surfaces vanishes. Thus, by equation \eqref{eq: current}, we conclude that
\begin{equation}\label{eq: QS in metric and coords}
    \left(\mu \rho \right)_\alpha = 0,
\end{equation}
or equivalently that $\partial B/\partial \alpha = 0$. This symmetry of the field strength is simply quasisymmetry. Indeed, we define the quasisymmetry vector as the covariant basis vector $\boldsymbol{u}=\partial_\alpha \boldsymbol{r} = \rho \boldsymbol{\hat{b}}$. Then, using the quasisymmetry definition of \cite{Rodriguez2020}, it follows that locally
\begin{subequations}\label{eqs: governing eqs}
\begin{align}
    \boldsymbol{u}\cdot \bnabla B &= 0,\label{eq: QSa}\\ 
    \boldsymbol{u}\times\boldsymbol{B} &= \bnabla \psi,\label{eq: QSb} \\
    \bnabla\cdot\boldsymbol{u} &= 0.\label{eq: QSc}
\end{align}
\end{subequations}
Equation \eqref{eq: QSa} follows from $B=1/\mu\rho$ and equation \eqref{eq: QS in metric and coords}. Equation \eqref{eq: QSb} follows from $\boldsymbol{u}=\rho \boldsymbol{\hat{b}}$, $\boldsymbol{B}=\boldsymbol{\hat{t}}/\mu\rho$, $\boldsymbol{\hat{b}}\times\boldsymbol{\hat{t}}=\boldsymbol{\hat{n}}$, and $\bnabla \psi = \boldsymbol{\hat{n}}/\mu$. Note that equation \eqref{eq: QSb} usually has an additional minus sign, but we define $\boldsymbol{u}$ such that $\rho >0$ and $(\boldsymbol{\hat{t}},\boldsymbol{\hat{n}},\boldsymbol{\hat{b}})$ forms a right-handed coordinate system. Equation \eqref{eq: QSc} follows from $\bnabla \cdot \boldsymbol{u} = \partial_\alpha (\mu \rho)/ \mu \rho=0$. It is notable that we did not have to assume quasisymmetry; it emerged naturally from assuming geodesic field lines and zero current normal to the surfaces and revealed itself in geodesic coordinates.

We now demonstrate why QP surfaces are natural choices for LQP surfaces. We prove that, given a toroidal surface with zero enclosed toroidal current and pointwise zero normal current, the field lines are geodesics on that surface if and only if the magnetic field strength on the surface has QP symmetry.

Firstly, we prove the forward direction: geodesic field lines imply QP. We have already shown that LQP surfaces are quasisymmetric, so they possess a symmetry vector $\boldsymbol{u}$ and a corresponding symmetry coordinate $\alpha$. On a toroidal surface where $B$ is not constant on a flux surface, the integral curves of $\boldsymbol{u}$ must close, according to equation \eqref{eq: QSa}. The integral of $\boldsymbol{B}$ around a closed $\boldsymbol{u}$ integral curve (denoted $\partial S$) must vanish because $\boldsymbol{u}\cdot\boldsymbol{B}=0$ by the choice of geodesic coordinates, so
\begin{equation}
        0=\oint_{\partial \mathcal{S}} d\alpha \, \frac{\partial \boldsymbol{r}}{\partial \alpha}\cdot \boldsymbol{B}= \iint_\mathcal{S} d\boldsymbol{S} \cdot \boldsymbol{J},
\end{equation}
where the second equality follows from Stokes' theorem.  If $\partial \mathcal{S}$ closed toroidally, then $\iint_S d\boldsymbol{S}\cdot \boldsymbol{J}$ is the external poloidal current, which does not vanish. If $\partial \mathcal{S}$ closed poloidally, then $\iint_S d\boldsymbol{S}\cdot \boldsymbol{J}$ is the enclosed toroidal current, which does vanish. So, $\partial S$ as a poloidal curve is a possible solution. As a final possibility, if $\partial \mathcal{S}$ closed helically, then $\iint_S d\boldsymbol{S}\cdot \boldsymbol{J}$ is a linear combination between external poloidal and enclosed toroidal currents, which also does not vanish. So, $\partial \mathcal{S}$ must close poloidally. This yields QP.

The poloidal current exterior to the flux surface does not vanish, whereas the toroidal current enclosed by the flux surface does vanish.

We now show the backward direction: QP implies geodesic field lines, given zero enclosed toroidal current and $\boldsymbol{J}\cdot\boldsymbol{\hat{n}}=0$. By quasisymmetry and the vanishing of the normal current, we have $(\eta/\mu \rho)_l=0$. For an irrational surface, this means that $\eta/\mu \rho$ is constant on the surface, since field lines get arbitrarily close to every point on the surface. For rational surfaces, assuming that the rotational transform is not constant and we have QP in an infinitesimal volume around the surface, we also have that $\eta/\mu \rho$ is constant by continuity. Then, we integrate the magnetic field along the poloidal symmetry curve $\partial \mathcal{S}$. Using that the enclosed toroidal current vanishes and Stokes' theorem, this integral vanishes:
\begin{equation}
        0=\oint_{\partial \mathcal{S}} d\alpha \, \frac{\partial \boldsymbol{r}}{\partial \alpha}\cdot \boldsymbol{B}= \oint_{\partial \mathcal{S}} d\alpha \, \frac{\eta}{\mu \rho}.
\end{equation}
As $\eta/\mu\rho$ is constant, this implies that $\eta=0$, and consequently that we have geodesic field lines. 

In appendix \ref{apdx: Boozer coords}, we make the connection with \cite{Boozer1981} coordinates, which can be defined since $\boldsymbol{J}\cdot \boldsymbol{\hat{n}}=0$ \citep{Rodriguez2021}. We also show that, under the further assumption of MHS force balance, the parallel current on a QP surface vanishes whenever the enclosed toroidal current vanishes. Thus, we conclude from this subsection that QP is the natural global analogue for LQP surfaces. However, LQP surfaces, defined locally, constitute a larger class of possible surfaces, toroidal or not (e.g., magnetic mirrors).

\subsection{The surface equations}\label{subsec: the surface equations}

In this subsection, we present a reduced, decoupled model for single LQP flux surfaces. We first discuss what it means to have defined a single LQP surface in the language of the moving frame.

A magnetic field is exactly described by $\eta,\rho,\lambda,\mu,\nu,\boldsymbol{K},\boldsymbol{\chi},$ and $\boldsymbol{\sigma}$ as functions of $(l,\alpha,\psi)$, provided that the geometric constraints of \S\ref{sec: moving frame representation} are obeyed. In contrast, a single flux surface, taken as an embedded surface in real space, is precisely described by $\eta,\rho,\boldsymbol{K},$ and $\boldsymbol{\chi}$ as functions of $(l,\alpha)$, provided that equations \eqref{eq: frame compatibility a} and \eqref{eqs: lb compatibility} are satisfied. This result of classical differential geometry is referred to as Bonnet's theorem (see, e.g., \oldS3.6 in \citealt{struik1950lectures}). A magnetic field requires additional information because it is a 3D bundle of field lines, whereas a single flux surface is a 2D manifold.

Even if we consider only a single LQP flux surface, finding an LQP surface is in principle a 3D problem. This is because defining the field strength and current on a flux surface requires information about field lines away from that surface. For the field strength, given $B= 1/(\mu \rho) = 1/(\boldsymbol{r}_l\cdot \boldsymbol{r}_\alpha \times \boldsymbol{r}_\psi)$, we need to know field lines in the infinitesimal neighbourhood of the surface, i.e., we need to know $\boldsymbol{r}(l,\alpha,\psi_0+\delta \psi)$ to first order in $\delta \psi$, where $\psi_0$ labels the relevant flux surface.

In general, describing currents on a flux surface requires information about field lines further from the surface, i.e. $\boldsymbol{r}(l,\alpha,\psi_0+\delta \psi)$ to two orders in $\delta \psi$. This is manifest in the expression \eqref{eq: current} for the current as terms like $\mu_\psi$ depend on $\boldsymbol{r}_{\psi \psi}$ (see equation \eqref{eqs: covariant basis vectors b}). However, when the field lines are geodesics on the surface, one requires information only about field lines in the immediate infinitesimal neighbourhood, i.e. to first order in $\delta \psi$, to define the currents normal to the surface and parallel to the field lines. This is because when $\eta=0$ on the flux surface, which is the geodesic condition in geodesic coordinates as described in \S\ref{subsec: relation to QP}, there are no terms involving $\boldsymbol{r}_{\psi\psi}$ contributing to the normal and parallel current, as can be seen from equation \eqref{eq: current}. 

Therefore, to define a single LQP surface, we need to define a particular $\boldsymbol{r}(l,\alpha,\psi_0+\delta\psi)$ to first order in $\delta \psi$. In other words, a single LQP surface is defined by field lines on and directly adjacent to that surface. Geometrically, defining such field lines is equivalent to providing $\lambda,\mu,\nu,$ and $\boldsymbol{\sigma}$ on the surface as functions of $(l,\alpha)$ at fixed $\psi=\psi_0$ and $\boldsymbol{K},\boldsymbol{\chi},\rho,$ and $\eta$ as functions of $(l,\alpha,\psi)$ for $\psi$ in the immediate neighbourhood of $\psi_0$, provided that equations \eqref{eq: frame compatibility}, \eqref{eqs: lpsi compatibility}, and \eqref{eqs: bpsi compatibility} hold at $\psi=\psi_0$ and equations \eqref{eqs: lb compatibility} hold for $\psi=\psi_0+\delta \psi$ for infinitesimal, arbitrary $\delta \psi$. Solving these equations for those variables merely provides the requisite structure; an LQP surface must also satisfy the physical constraints of having (i) geodesic field lines, (ii) zero normal current, and (iii) zero parallel current:
\begin{subequations}\label{eqs: physical constraints}
    \begin{align}
        \eta &= 0 \quad \text{for } \psi=\psi_0+\delta\psi,\label{eqs: physical constraints b}\\
        (\mu \rho)_\alpha &= 0\quad \text{for } \psi=\psi_0,\label{eqs: physical constraints c}\\
        \lambda_\alpha &= 0 \quad \text{for } \psi=\psi_0.\label{eq: assumption parallel current weak} 
    \end{align}
\end{subequations}
The first equation corresponds to having geodesic field lines and choosing $(l,\alpha)$ to be geodesic coordinates (as discussed in \S\ref{subsec: relation to QP}). We do not just impose $\eta=0$ on the relevant flux surface, but also in the neighbourhood of the flux surface. This reflects the requirement that a finite volume of surfaces possess the physical benefits of geodesic field lines. The second equation corresponds to zero normal current and is equivalent to quasisymmetry (also discussed in \S\ref{subsec: relation to QP}). Given the previous two equations, the third equation corresponds to zero parallel current on the relevant flux surface. 

We note one crucial simplification arising from field lines being geodesics. By equation \eqref{eq: lb compatibility a}, we deduce that $K_2=0$. Thus, the Darboux frame on the surface reduces to the Frenet-Serret frame for the field lines, since equation \eqref{eqs: defs for K, chi, sigma a} becomes the Frenet-Serret formulae \citep{Serret1851,Frenet1852}. To comply with the usual notation, we replace `$\boldsymbol{K}$' with the curvature $\kappa \coloneq -K_3$ (equal to the normal curvature) and torsion $\tau \coloneq -K_1$. We note that the curvature defined in this way can be negative, precisely when the principal normal of the field lines is anti-aligned with the surface normal.

We have determined the relevant variables and equations for finding a single LQP surface using the moving frame formalism. In appendix \ref{apdx: derivation of surface equations}, we manipulate and reduce these equations to obtain the main result of this paper: the geometry of a single LQP flux surface is determined by the solution to the \textit{surface equations}
\begin{subequations}\label{eqs: surface equations}
    \begin{align}
        \kappa_\alpha &= -2\rho_l \tau - \rho \tau_l, \label{eq: codazzi-mainardi a}\\
        \tau_\alpha &= \rho_l \kappa + \left( \frac{\rho_{ll}-\rho \tau^2}{\kappa} \right)_l,\label{eq: codazzi-mainardi b}\\
        \left( \frac{\kappa}{\rho}\right)_\alpha &= 0.\label{eq: rho_div_kappa_constraint}
    \end{align}
\end{subequations}
For the reader who skipped to these equations, we reiterate that $\kappa$ is the signed curvature of the field lines, $\tau$ is the torsion of the field lines, $(l,\alpha)$ are a particular choice of field-line-following coordinates where $\partial_\alpha \boldsymbol{r}=\boldsymbol{u}$ is the quasisymmetry vector, and $\rho=|\boldsymbol{u}|$ is the `orthogonal distance' between the field lines. 

The surface equations have three remarkable properties. Firstly, the surface equations depend only on the geometry of the flux surface and the field lines on that flux surface (through the variables $(\rho,\kappa,\tau)$); they do not depend on field lines away from the flux surface. How can this be if the field strength and currents on the flux surface depend on field lines away from the flux surface? This is because of a second remarkable property: given a solution to the surface equations (representative of an embedded surface foliated by field lines), there exist (at least locally) some neighbouring field lines that would confer upon the surface the relevant LQP properties of zero normal and parallel currents. We show this mathematically by providing a set of equations that (along with the surface equations) are equivalent to the original problem for finding an LQP surface, and then showing that these equations admit solution. These equations are derived alongside the surface equations in appendix \ref{apdx: derivation of surface equations} and are presented here:
\begingroup
\allowdisplaybreaks
\begin{subequations}\label{eqs: complete surface solution}
    \begin{align}
        \eta &= 0,\label{eq: complete surface sln eta}\\
        K_2 &=0,\\
        \chi_1 &= -\frac{\rho_{ll}-\rho \tau^2 }{\kappa},\label{eqs: complete surface solution chi1}\\
        \chi_2 &= \rho_l,\label{eqs: complete surface solution chi2}\\
        \chi_3 &= \rho \tau,\label{eqs: complete surface solution chi3}\\
        \lambda_\alpha &= 0,\label{eqs: complete surface solution 0}\\ 
        \lambda_l &= \kappa \mu, \label{eqs: complete surface solution a}\\
        \nu_l &= \frac{\rho_l}{\rho}\nu - 2 \tau \mu,\label{eqs: complete surface solution b}\\
        \mu_\alpha &= - \frac{\rho_\alpha}{\rho}\mu,\label{eqs: complete surface solution c}\\
        \sigma_1 &= - \frac{\rho_\alpha}{\rho^2} \mu - \tau \lambda - \frac{\nu}{\rho}\frac{\rho_{ll}-\rho \tau^2}{\kappa},\label{eqs: complete surface solution c1}\\
        \sigma_2 &= \frac{\rho_l}{\rho}\nu -\tau \mu,\label{eqs: complete surface solution c2}\\
        \sigma_3 &= -\mu_l -\kappa \lambda + \tau \nu,\label{eqs: complete surface solution c3}\\
        \rho_\psi &= \nu_\alpha + \rho_{l} \lambda + \frac{\rho_{ll}-\rho \tau^2}{\kappa} \mu,\label{eqs: complete surface solution e}\\
        \eta_\psi &= 0,\\    
        \kappa_\psi &= \mu_{ll} + (\kappa^2 + 3 \tau^2)\mu + \kappa_l \lambda + \frac{\kappa_\alpha}{\rho}\nu,\label{eqs: complete surface solution d}\\
        \tau_\psi &= \frac{\tau_\alpha}{\rho}\nu + \tau_l \lambda + \frac{\rho_\alpha}{\rho^2}\mu_l + \left( \left( \frac{\rho_\alpha}{\rho^2}\right)_l - 2 \frac{\tau}{\rho} \frac{\rho_{ll}-\rho\tau^2}{\kappa} \right)\mu,\label{eqs: complete surface solution f}\\
        K_{2\psi} &= 0,\\
        \chi_{1\psi} &= -\left(\frac{\rho_{ll}-\rho \tau^2}{\kappa} \right)_\psi, \\
        \chi_{2\psi} &= \rho_{l \psi}, \\
        \chi_{3 \psi} &= (\rho \tau)_\psi.\label{eq: complete surface sln chi3psi}
    \end{align}
\end{subequations} 
\endgroup
These equations apply at fixed $\psi=\psi_0$. Taken together with the surface equations, they enforce the relevant physical and geometric conditions. We now justify why these equations are solvable, given a solution $(\rho,\kappa,\tau)$ to the surface equations. First, we consider the variables $(\lambda,\mu,\nu)$. They are determined by the homogeneous, linear differential equations \eqref{eqs: complete surface solution 0}--\eqref{eqs: complete surface solution c}. By the \cite{Frobenius+1877+230+315} theorem, there exists a solution $(\lambda,\mu,\nu)$ to these differential equations. This is because equations \eqref{eqs: complete surface solution 0} and \eqref{eqs: complete surface solution a} are compatible: both equations agree that $\lambda_{\alpha l}=0$, since $(\kappa \mu)_\alpha=0$ by the surface equation \eqref{eq: rho_div_kappa_constraint} and the quasisymmetry condition, presented here as equation \eqref{eqs: complete surface solution c}. To uniquely specify a solution to these equations, certain boundary conditions must be specified: $\mu$ as a function of $l$ at fixed $\alpha$ (which corresponds to choosing the field strength, see \S\ref{subsec: what is special}), $\nu$ as a function of $\alpha$ at fixed $l$, and $\lambda$ at a particular point. There is less freedom to specify boundary conditions in $\lambda$ because it is controlled by two equations, \eqref{eqs: complete surface solution 0} and \eqref{eqs: complete surface solution a}. The remaining functions $(\eta,K_2,\boldsymbol{\chi},\boldsymbol{\sigma},\rho_\psi,\eta_\psi,\tau_\psi,\boldsymbol{K}_\psi,\boldsymbol{\chi}_\psi)$ are algebraically and compatibly determined in terms of $\rho,\kappa,\tau,\lambda,\mu$, and $\nu$ by equations \eqref{eq: complete surface sln eta}--\eqref{eqs: complete surface solution chi3} and \eqref{eqs: complete surface solution c1}--\eqref{eq: complete surface sln chi3psi}.

A solution to these equations, alongside a solution to the surface equations, represents a single LQP flux surface. This is because, as discussed, a single LQP surface is defined by field lines in an infinitesimal volume around that surface, or equivalently, by the functions $\lambda,\mu,\nu,\sigma,\boldsymbol{K}, \boldsymbol{\chi},\rho,\eta,\boldsymbol{K}_\psi, \boldsymbol{\chi}_\psi,\rho_\psi,$ and $\eta_\psi$ at fixed $\psi=\psi_0$ as functions of $(l,\alpha)$. Thus, the problem of finding an LQP flux surface has been reduced to solving the surface equations, since the remaining part of the problem is solvable given a solution to the surface equations. It is surprising that we do not have an overdetermined system, despite the numerous equations involved. We discuss the physical reasons for this in \S\ref{subsec: what is special}.

A third and final remarkable property of the reduction to the surface equations is that $(\rho,\kappa,\tau)$ \textit{completely determine} the surface and the field lines on it. Intuitively, this is clear: $(\kappa,\tau)$ determine the field lines individually, and $\rho$ tells us the spacing between the field lines. Formally, we can recall Bonnet's theorem (stated at the start of this subsection) and note that $\boldsymbol{\chi}$ is algebraically determined by $(\rho,\kappa,\tau)$ through equations \eqref{eqs: complete surface solution chi1}--\eqref{eqs: complete surface solution chi3}. The relevant compatibility conditions \eqref{eq: frame compatibility a} and \eqref{eqs: lb compatibility} are satisfied identically when $\boldsymbol{\chi}$ is expressed in this way, with the exception of two components of \eqref{eq: frame compatibility a}, which are equivalent to the first two surface equations. Thus, the first two surface equations are purely geometric conditions, as viewed by geodesics. Curiously, the physics of vanishing normal and parallel current reduces to the simple equation \eqref{eq: rho_div_kappa_constraint}. 

To summarise this subsection, the problem of finding an LQP surface has been reduced from a 3D problem to a 2D problem, since we need only solve the surface equations for the geometry of an embedded surface. An LQP surface is defined by its properties of field strength and currents, which depend on field lines away from the surface. However, once we have a solution to the surface equations, we can locally guarantee that there exist neighbouring field lines that would give that surface the relevant properties of an LQP flux surface, viz. zero parallel and normal currents. We can make this guarantee because the variables that define the neighbouring field lines can be found by solving a well-posed set of equations. In this sense, the surface equations are decoupled and complete. Appendix \ref{apdx: classical differential geometry} makes a connection with the language of classical differential geometry, for the reader who is fluent in such a language.

Finally, we comment on magnetohydrostatic (MHS) force balance \eqref{eq: MHS force balance}, which depends on the current in the binormal direction (perpendicular to the field lines and tangential to the flux surface). This component of current has not yet been considered and need not vanish, unlike the parallel and normal components of current. By substituting the expression \eqref{eq: current} for the current into MHS force balance \eqref{eq: MHS force balance}, we find
\begin{equation}\label{eq: complete surface sln with MHS}
    \mu_\psi = -\frac{\rho_\psi}{\rho}\mu + p'(\psi) \rho^2 \mu^3 - \rho \mu^2 \left( \frac{\lambda}{\mu \rho}\right)_l.
\end{equation}
Because we look for a solution on a single surface, \eqref{eq: complete surface sln with MHS} does not lead to overdetermination, since $\mu_\psi$ was unconstrained by the system of equations in \eqref{eqs: complete surface solution}. Physically, this is because the current in the binormal direction is affected by field lines even further from the surface. Formally, the presence of $\mu_\psi$ in the expression for this component of the current implies that one requires $\boldsymbol{r}(l,\alpha,\psi_0+\delta \psi)$ to two orders in $\delta \psi$ to define $\boldsymbol{J}\cdot \boldsymbol{\hat{b}}$ on the flux surface $\psi=\psi_0$. For a single LQP flux surface, these `far away' field lines do not affect the parallel or normal currents, so we can choose them to satisfy MHS force balance on the relevant surface: hence, there is no overdetermination by imposing force balance when considering a single LQP flux surface. This is a fortunate and unique scenario: typically in studies of MHS equilibria, imposing MHS force balance on a quasisymmetric equilibrium creates overdetermination \citep{garren_existence_1991,sengupta2024periodickortewegdevriessoliton}. Moreover, equation \eqref{eq: complete surface sln with MHS} implies that $(\mu \rho)_{\psi \alpha}=0$, which is just a statement of quasisymmetry in the neighbourhood of the surface. Therefore, all considerations thus far, including the surface equations, are consistent with a finite-beta equilibrium. However, it must be cautioned that equation \eqref{eq: complete surface sln with MHS} does determine the allowable size of $\delta \psi$ for which the present description is accurate. 

\subsection{A sketch for solving the surface equations through optimisation}\label{subsec: properties of surface equations}
In this subsection, we sketch how the surface equations could be solved as an optimisation problem for generating LQP flux surfaces. This could be used as part of a larger optimisation for generating a practical QP stellarator field or for quickly and efficiently generating flux surfaces to study trends in QP equilibria. This approach would be intrinsically faster than usual MHD equilibrium solvers (e.g., {\tt VMEC}, \citealt{Hirshman1983}, or {\tt DESC}, \citealt{Dudt2020}) because we do not need to directly solve for MHD equilibrium within a 3D volume. For concreteness, we take the example of a toroidal QP flux surface, but a similar process could be designed for the generation of asymmetric magnetic mirror flux surfaces with the desirable properties of an LQP surface.

Computationally, we work in the coordinate system $(\vartheta,\varphi)$, where $\varphi$ is some particular toroidal angle coordinate and $\vartheta$ is some particular poloidal angle coordinate. The flux surface is parameterised by the single-valued functions $R(\vartheta,\varphi)$ and $Z(\vartheta,\varphi)$, where $R$ and $Z$ correspond to the usual cylindrical coordinates with respect to an axis passing through the centre of the torus. In practice, one could work in discrete Fourier modes of $(\vartheta,\varphi)$, e.g., $R_{mn}$, up to some chosen level of precision. 

On the toroidal surface, we also parameterise a foliation of closed poloidal loops. In an exact QP solution, these would correspond to lines of constant $l$ (in the geodesic coordinates used throughout this paper). Equivalently, in an exact QP solution, they are the field lines for the field $\boldsymbol{u}=\partial_\alpha \boldsymbol{r}$, or the loops along which the magnetic field strength would be constant (though we do not solve for the magnetic field strength in this 2D problem). We take the poloidal loops to be level sets of some function $\Lambda(\vartheta,\varphi)$, which is necessarily periodic in $\vartheta$. The field lines are taken as perpendicular to these poloidal loops. For specificity, we could take $\Lambda$ to be the distance along a chosen field line from some reference poloidal loop. In the solved problem where the field lines are geodesics, this $\Lambda$ becomes independent of the chosen field line and is equivalent to our choice of arc length coordinate $l$.

We generate LQP flux surfaces by solving the surface equations as an optimisation problem. The free parameters are the surface, through $R(\vartheta,\varphi)$ and $Z(\vartheta,\varphi)$, and the poloidal loops, through the function $\Lambda(\vartheta,\varphi)$. The function $\Lambda(\vartheta,\varphi)$ is constrained to have level sets that close poloidally. The cost function, $f[\Lambda,R,Z]$, has two parts, $f_1[\Lambda,R,Z]$ and $f_2[\Lambda,R,Z]$, where $f_1[\Lambda,R,Z]=0$ ensures that the field lines are geodesics and if $f_2[\Lambda,R,Z]=0$ also, the third surface equation, \eqref{eq: rho_div_kappa_constraint}, is solved. We now consider these two parts in more detail.

Firstly, we enforce that the field lines are geodesics. The tangent vector to the field lines $\boldsymbol{\hat{t}}$ is obtained by calculating the orthogonal to the poloidally closed level sets of $\Lambda$, a calculation which depends on $R$ and $Z$. Then, using the tangent vector field $\boldsymbol{\hat{t}}$, the geodesic curvature can be calculated, in a way that likewise depends on $R$ and $Z$ and is also calculable via well-established methods. The field lines are geodesics if and only if the geodesic curvature is zero everywhere, so we take this part of the cost function to be
\begin{equation}
    f_1[\Lambda,R,Z] = \int_0^{2 \pi} d\vartheta \int_0^{2 \pi} d\varphi \,\sqrt{g}_\text{surf} \, |\kappa_g|,
\end{equation}
where $\kappa_g \equiv K_2$ is the geodesic curvature and the surface Jacobian $\sqrt{g}_\text{surf}$ ensures the integral is a surface average.

The second part of the cost function is connected with the third surface equation \eqref{eq: rho_div_kappa_constraint} and is somewhat more subtle. We define the field line label $\tilde{\alpha}$ to be constant along the magnetic field lines and go from $0$ to $2 \pi$ (for specificity) along the closed poloidal loops. We then take $s$ to be the arc length along the closed poloidal loop and define $\tilde{\rho} \coloneq \partial s/\partial \tilde{\alpha}$. For the third surface equation, we must also calculate the normal curvature $\kappa_n \equiv -K_3$ from $\boldsymbol{\hat{t}}$. When the geodesic curvature is zero, the normal curvature is the same as the curvature $\kappa$.

We take the second part of the cost function to be
\begin{equation}\label{eq: f2}
    f_2[\Lambda,R,Z] = \int_0^{2 \pi} d\vartheta \int_0^{2 \pi} d\varphi \,\sqrt{g}_\text{surf}\, \left|\left( \ln \frac{\kappa_n}{\tilde{\rho}}\right)_{l \tilde{\alpha}}\right|.
\end{equation}
To justify this expression, we show that precisely when
\begin{equation}\label{eq: verification version of rho_div_kappa}
    \left( \ln \frac{\kappa}{\tilde{\rho}}\right)_{l \tilde{\alpha}}=0,
\end{equation}
there exists a transformation $\alpha=f(\tilde{\alpha})$ such that the third surface equation \eqref{eq: rho_div_kappa_constraint} is satisfied. To show this, we suppose equation \eqref{eq: verification version of rho_div_kappa} is true. Then, 
\begin{equation}\label{eq: eq defining g alphatilde}
    \left( \ln \frac{\kappa}{\tilde{\rho}}\right)_{ \tilde{\alpha}}=g(\tilde{\alpha}),
\end{equation}
for some particular $g$. We define $\alpha = f(\tilde{\alpha})$. It follows that $\rho = \tilde{\rho} \frac{df}{d \tilde{\alpha}}$. We choose $f$ to be
\begin{equation}
    f(\tilde{\alpha})= \int_0^{\tilde{\alpha}} da \, \exp\left( \int_0^a db \, g(b)\right).
\end{equation}
$f$ has all the necessary features of a transform between periodic coordinates. First, it is monotonic. Secondly, for $\tilde{\alpha}$ increasing by a period, which is $2\pi$ here for specificity, $f$ increases by a fixed amount. This relies on $\int_0^{2\pi} d \tilde{\alpha} \; g(\tilde{\alpha})=0$, which follows from equation \eqref{eq: eq defining g alphatilde} and the fact that $\kappa$ and $\tilde{\rho}$ are single-valued. Since $f$ solves the differential equation
\begin{equation}
    \frac{d^2 f}{d \tilde{\alpha}^2} =  \frac{df}{d \tilde{\alpha}} g(\tilde{\alpha}),
\end{equation}
the third surface equation \eqref{eq: rho_div_kappa_constraint} is satisfied. In the 2D picture, there is a `gauge' freedom $\alpha \rightarrow f(\alpha)$ (corresponding to $\rho \rightarrow \rho/f'(\alpha)$) in the orthogonal coordinate $\alpha$, but the third surface equation \eqref{eq: rho_div_kappa_constraint} still imposes a non-trivial constraint, through equation \eqref{eq: verification version of rho_div_kappa}.

We do not need to verify the first two surface equations, \eqref{eq: codazzi-mainardi a} and \eqref{eq: codazzi-mainardi b}. These are identically satisfied when the field lines are geodesics, with $\kappa=\kappa_n$ and $\rho\rightarrow \tilde{\rho}$. Indeed, these two surface equations are purely geometric constraints and are satisfied by any geodesic foliation of an embedded surface. In this setup, the geometry is encoded through $R(\vartheta,\varphi)$ and $Z(\vartheta,\varphi)$.

We also note that in practice, solving for $f_2[\Lambda,R,Z]=0$ makes sense only if the field lines are sufficiently close to geodesics (i.e., $\kappa_n \gg \kappa_g$). In a computational optimisation problem, this could be achieved by including the $f_2$ part only when $f_1$ is sufficiently small. Moreover, there could be numerical issues with the logarithm in equation \eqref{eq: f2}. If geodesic curvature is close to zero everywhere (i.e. $f_1$ is sufficiently small), then normal curvature is likely to only be zero on isolated points or curves, lest the surface have flat regions. In any case, the integrand can be transformed to regularise these possible issues, e.g. by using $\left|\left( \tanh\left(\ln \frac{\kappa_n}{\tilde{\rho}}\right)\right)_{l \tilde{\alpha}}\right|$ instead.

In practice, once a toroidal surface has been selected by this method, it can be used as the boundary for a complete equilibrium solver. In vacuum, the boundary shape completely determines the magnetic field inside the boundary. However, because of the locality of the assumptions behind LQP surfaces, there is no guarantee that there are nested QP flux surfaces within the boundary of a surface selected by this method. In this case, a second optimisation, using the equilibrium solver, could be used to fine-tune the boundary surface to give the desired global properties. This approach has reasons to be better than existing methods for selecting initial conditions for QP. For instance, a heuristic, near-axis construction (such as the one devised by \cite{Goodman23} and applied to QP in \cite{haque2026}) necessarily contains deviations from exact QP, since QP is not possible through the near-axis expansion. The current framework is not limited to near-axis. Nonetheless, it remains as future work to verify whether this approach can indeed lead to improved QP quality. 

\subsection{What is so special about LQP surfaces?}\label{subsec: what is special}
There are two unique features of LQP surfaces that lead to the drastic simplification of the surface equations. Indeed, such a reduction cannot be found generally. If we were to impose only quasisymmetry as a physical constraint (i.e., only equation \eqref{eqs: physical constraints c}), the moving frame formalism would lead to equations that are highly coupled and overdetermined. In such cases, describing the magnetic field through a moving frame is not particularly useful.

The first special feature is that, for LQP surfaces, the surface geometry is decoupled from the magnetic field strength. In the case of general quasisymmetry with zero normal current, $\boldsymbol{u}\cdot\boldsymbol{B}=\eta/\mu\rho$ is constant on a flux surface \citep{Rodriguez2020}. This couples a quantity that influences the geometry of the flux surface, $\eta$, with a quantity that is determined by field lines in the neighbourhood of the surface, namely the field strength $1/\mu\rho$. LQP surfaces avoid this coupling by having geodesic field lines, which implies $\eta=0$ everywhere. Notably, given a single LQP surface, it is always possible to construct another LQP surface with the same flux surface shape but a different magnetic field strength profile (so long as $\partial B/\partial \alpha=0)$. This is because we can take a complete solution to the LQP equations \eqref{eqs: surface equations} and \eqref{eqs: complete surface solution}, and construct another solution through $\mu \rightarrow g(l) \mu$ for arbitrary function $g$.

There is a second way in which LQP surfaces are special. Without assuming geodesic field lines, the general expression for the parallel current in the moving frame representation is
\begin{equation}
    J_\parallel = \frac{1}{\mu \rho}  \left(    \left( \frac{\eta}{\mu \rho} \right)_\psi -\left( \frac{\lambda}{\mu \rho} \right)_\alpha\right) + \frac{\eta}{\mu \rho}  \left(   \left( \frac{\lambda}{\mu \rho} \right)_l -\left( \frac{1}{\mu \rho} \right)_\psi \right),
\end{equation}
where we have continued to assume that the normal current vanishes. As an aside, the second set of terms is the Pfirsch-Schl\"uter current, which vanishes for geodesic field lines ($\eta=0$). For $\eta=0$, the expression for the parallel current ceases to depend on $\mu_\psi$. This means that for the particular case of geodesic field lines, the parallel current is determined only by field lines in the immediate neighbourhood of the surface, not by field lines at second order from the surface. Despite being more restrictive---there are fewer field lines that can contribute to having zero parallel current---this fact leads to mathematical simplification, since the equations are decoupled from field lines that are not in the immediate neighbourhood of the surface. In other words, in the general non-geodesic case, we have to consider a `thicker' volume around the surface in order to evaluate the parallel current, which means there are more variables coupled into the equations (though optimistically, this also represents increased freedom). It is also this fact that means our single LQP surface can always be force-balanced without additional overdetermination, since the binormal current $\boldsymbol{J}\cdot \boldsymbol{\hat{b}}$ can be manipulated without affecting the parallel current, by altering field lines further from the surface through $\mu_\psi$, in accordance with equation \eqref{eq: complete surface sln with MHS}.

\section{Two fundamental classes of solutions to the surface equations}\label{sec: possibility of irrational QP surfaces}
We have thus far considered only local details of the magnetic field, except in \S\ref{subsec: properties of surface equations}, where we used the surface equations to define a reduced optimisation problem for finding toroidal QP flux surfaces. Here, we analytically apply global considerations to the surface equations. In this section, we identify two natural classes of solutions to the surface equations: the first contains `magnetic mirrors' as a solution, and the second has attractive and analytically tractable field strength properties.

We consider searching for toroidal, QP solutions. A toroidal QP surface is foliated by closed poloidal loops. Along these loops, the magnetic field strength is constant, the covariant basis vector $\partial_\alpha \boldsymbol{r}$ (also the quasisymmetry vector) is tangent, and the field lines are orthogonal. In the $(l,\alpha)$ coordinate system, these loops are given by $l=\text{const}$. (In \S\ref{subsec: properties of surface equations}, these poloidal loops were part of the free parameters for the optimisation problem.) We take $L$ to be the distance along a field line until it returns to the same poloidal loop, this value being independent of the chosen field line. Equivalently, $L$ is the distance along a field line between $\phi=0$ and $\phi=2 \pi$ for Boozer angle $\phi$, as defined in appendix \ref{apdx: Boozer coords}.

Thus, for a toroidal QP surface, $\rho$, $\kappa$, and $\tau$ are periodic in $\alpha$, but are generally aperiodic in $l$, given a field line that doesn't close on itself. These functions are single-valued in real space, so they are also constrained by the fact that  $(l+L,\alpha)$ and $(l,\alpha + 2\pi \iota)$ are the same point on the surface, where we define the rotational transform $\iota$. Unfortunately, beyond these necessary features, there are no \textit{a priori} constraints on $\rho$, $\kappa$, and $\tau$ that would ensure a toroidal surface. Thus, one either has to impose a numerical optimization scheme that enforces toroidicity, such as that described in \S\ref{subsec: properties of surface equations}, or check \textit{post facto} whether a solution to the surface equations is a toroidal surface. 

We now proceed in attempting to find analytic, toroidal QP solutions. From the first and third surface equations \eqref{eq: codazzi-mainardi a} and \eqref{eq: rho_div_kappa_constraint}, we have the `conservation law'
\begin{equation}
    \frac{\partial}{\partial \alpha} \left( \frac{\rho \kappa}{2}\right) +\frac{\partial}{\partial l} \left( \rho^2 \tau\right)=0.
\end{equation}
We integrate this equation around one of the closed poloidal loops, which gives
\begin{equation}
    \oint d\alpha \, \rho^2 \tau = C,
\end{equation}
where $C$ is some constant, and we have used the fact that $(\rho,\kappa,\tau)$ are single-valued. We define $\gamma \coloneq \kappa/\rho$, which is independent of $\alpha$ by the third surface equation \eqref{eq: rho_div_kappa_constraint}. Then,
\begin{equation}\label{eq: thing for periodicity}
    \oint d\alpha \, \kappa^2 \tau = C\gamma^2.
\end{equation}
The left-hand side of this equation is invariant under $l \rightarrow l+L$, since $\kappa$ and $\tau$ are single-valued, so $\gamma(l)$ is periodic with period $L$. This is evidently true for an irrational surface, which is where $\iota$ is an irrational number. This is because $(l+L,\alpha)$ and $(l,\alpha + 2\pi \iota)$ are the same point on the surface, $\gamma$ is smooth and independent of $\alpha$, and positive integer multiples of an irrational, taken modulo $2\pi$, are dense in $[0,2\pi]$.

The fact that $\gamma$ is periodic and $(\rho,\kappa,\tau)$ are generally aperiodic in $l$ couples periodic and aperiodic terms in the first two surface equations \eqref{eq: codazzi-mainardi a} and \eqref{eq: codazzi-mainardi b}. This complicates the route to a global, toroidal solution, but a separation of variables can be achieved by substituting $\kappa = \rho\gamma$. Then, we have
\begin{subequations}
    \begin{align}
        \gamma &= -\frac{2\rho_l \tau + \rho \tau_l}{\rho_\alpha},\\
        \gamma_l &= \frac{K+\tau^2}{\gamma \rho_l \rho-\tau_\alpha -\frac{ K + \tau^2}{\gamma}},
    \end{align}
\end{subequations}
where $K \coloneq -\rho_{ll}/\rho$ is the Gaussian curvature. Solving these two equations analytically in complete generality is not feasible, so we consider some natural classes of solutions. Given that $\gamma$ is periodic in $l$, two  natural classes emerge: a) $(\rho,\kappa,\tau)$ are independent of $\alpha$, and b) $\gamma$ is constant. 

\begin{figure}
    \centering
    \includegraphics[width=0.7\linewidth]{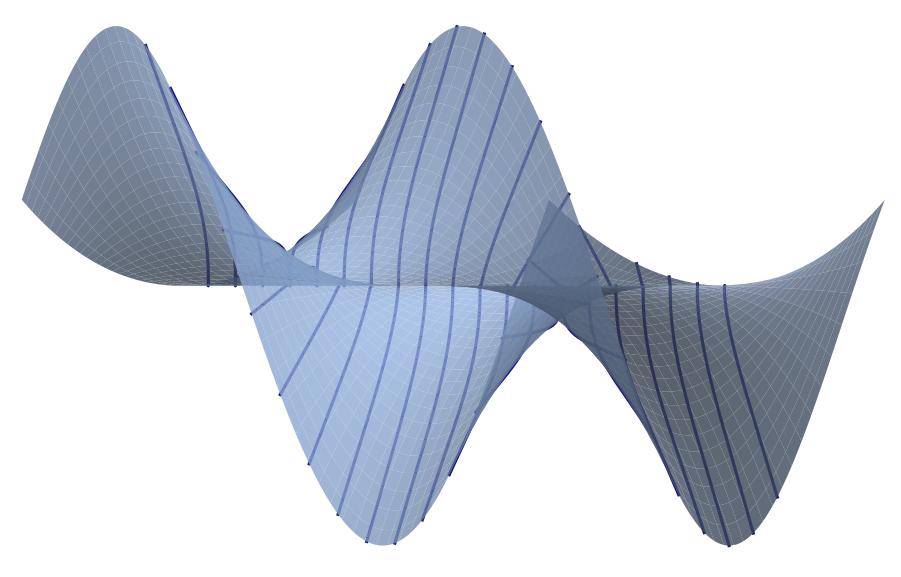}
    \caption{An example of a generalised helicoid, with $\boldsymbol{r}= \left(-0.6 u^2 \cos b,-0.6 u^2 \sin b, u+ b \right)$ for $u \in [-2,2]$ and $b \in [0,2\pi]$. The magnetic field lines (shown in dark blue) are geodesics orthogonal to the helices given by $u=\text{const}$. It should be cautioned that this surface is not poloidally closed, so it is not a fruitful solution for investigating QP. }
    \label{fig:generalized helicoid}
\end{figure}

\begin{figure}
    \centering
    \includegraphics[width=0.7\linewidth]{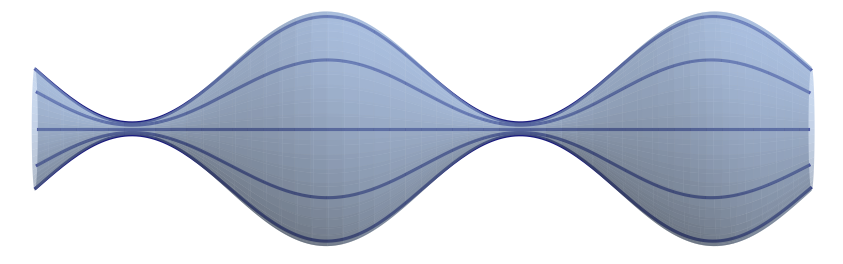}
    \caption{An example of a surface of revolution, with $\boldsymbol{r}= \left((1+ 0.9 \sin u)  \cos b,(1+ 0.9 \sin u) \sin b, u \right)$ for $u \in [-2 \pi,2 \pi]$ and $b \in [0,2\pi]$. The field lines are the meridians, highlighted in dark blue.}
    \label{fig:surf_of_rev}
\end{figure}

Case a) corresponds to generalised helicoids, as proved in appendix \ref{apdx: generalized helicoids}. An example is illustrated in figure \ref{fig:generalized helicoid}. In the further special case where $\tau=0$, we either have surfaces of revolution with the meridians as field lines (magnetic mirrors), as illustrated in figure \ref{fig:surf_of_rev}, or the surface traced out by a planar curve translated in the direction of the curve's binormal (i.e., surfaces of the form $z=f(x)$ with field lines given by $y=\text{const.}$). These two cases are the only LQP surfaces where the torsion of the field lines is zero, even if we were to allow for $\alpha$-dependence. Notably, all LQP surfaces where $\boldsymbol{u}$ can be extended to a Killing vector of 3D Euclidean space reduce to case a). This is also shown in appendix \ref{apdx: generalized helicoids}. 

Thus, case a) does not include any toroidal solutions. Nonetheless, in \S\ref{sec: numerical verification}, we see that numerically generated QP equilibria have sections that resemble magnetic mirrors, and in appendix \ref{apdx: perturbations to surfs of rev}, we perturb around surfaces of revolution to qualitatively explain some properties of the toroidal equilibria.

We now consider case b). In this case, $\rho \propto \kappa$. We have the same physical situation under the simultaneous rescaling $\alpha \rightarrow C \alpha$ and $\psi \rightarrow \psi/C$ for arbitrary constant $C$, so we can set $\gamma=1$ and $\rho=\kappa$. With $\rho=\kappa$, the first two surface equations \eqref{eq: codazzi-mainardi a} and \eqref{eq: codazzi-mainardi b} reduce to the Da Rios-Betchov equations \citep{da_rios_sul_1906,betchov_curvature_1965}, whilst the third equation \eqref{eq: rho_div_kappa_constraint} becomes trivial. Equivalently, using the definition of $\rho$, we can model the field lines as obeying the vortex filament equation (also known as the localised induction approximation),
\begin{align}
    \frac{\partial \boldsymbol{r}}{\partial \alpha} &= \kappa \boldsymbol{\hat{b}} \nonumber\\ &= \frac{\partial \boldsymbol{r}}{\partial l} \times \frac{\partial^2 \boldsymbol{r}}{\partial l^2},\label{eq: VFE b}
\end{align}
where the second line assumes that the principal normal and surface normal are in the same direction. If they are opposed, we still have the same physical equation, but under a `time-reversal' $\alpha \rightarrow - \alpha$. Surfaces traced out by a vortex filament are known as \textit{Hasimoto surfaces}, and we appropriately adopt this term for describing LQP surfaces where $\rho \propto \kappa$. Under the Hasimoto (\citeyear{hasimoto_soliton_1972}) map, the Da Rios-Betchov equations are equivalent to the \textit{nonlinear Schr\"odinger equation}.

The authors are aware of one known class of irrational, toroidal, Hasimoto surfaces. This solution already has prominence in an analytic MHS equilibrium, viz. the \cite{palumbo1968_Closed_MHD} configuration. The Palumbo configuration (or class of configurations) is an axisymmetric torus, made up of nested toroidal Hasimoto flux surfaces, with constant field strength on each surface. Remarkably, the Palumbo configuration is exact for an entire volume (not just a surface) and obeys the MHS force balance throughout for an arbitrary pressure profile $p(\psi)$. This series of toroidal Hasimoto surfaces is given by the periodic travelling wave solutions to the Da Rios-Betchov equations, $\kappa=\kappa(l+c\alpha)$ and $\tau = \tau(l+c\alpha)$, as identified by \cite{kida_vortex_1981}. The equation for such surfaces is given in an MHS context by equation (5.26) in \cite{SCHIEF_2003} and the surfaces are plotted in figure 1 therein. Unfortunately for our purposes, such surfaces, even when taken individually, cannot be QP, since the orthogonals to the geodesics wind helically and generally do not close. However, there may exist other toroidal Hasimoto surfaces that are consistent with QP, at least to an acceptable level of accuracy. In such a case, the magnetic field strength presents desirable features, as we show in the next subsection.

\subsection{The magnetic field strength on toroidal Hasimoto surfaces}\label{subsec: field strength on Hasimoto}

We have thus far ignored the magnetic field strength profile. This was because, as discussed in \S\ref{subsec: what is special}, the question of field strength decouples from the question of surface geometry. However, with some additional weak assumptions, we can make further statements about the magnetic field strength on Hasimoto surfaces. We first establish a connection between Hasimoto surfaces and an attractive property for practical stellarators, termed the flat mirror property. We then derive an exact governing differential equation that determines the magnetic field strength profile in terms of the surface geometry quantities, $(\rho,\kappa,\tau)$.

Firstly, we show that a Hasimoto surface in vacuum is a \textit{flat mirror}, as defined and studied in \cite{velasco2023robust_flat_mirror}. Velasco \textit{et al.} show that such magnetic fields exhibit small radial transport of energy and good confinement of bulk and fast ions. A flat mirror is defined as a QI magnetic field satisfying
\begin{equation}\label{eq: flat mirror condition}
    \partial_\psi \left( B_\text{max} -B_{00}\right)=0 ,
\end{equation}
where $B_{00}=\iint d\theta d\phi \, B/4\pi^2$, with $(\theta, \phi$) as Boozer angles, defined in appendix \ref{apdx: Boozer coords}, and $B_\text{max}$ is the maximum of the magnetic field strength on the surface. 

In vacuum, we have $(\lambda/\mu\rho)_l = (1/\mu\rho)_\psi$, as can be seen from the expression for the current \eqref{eq: current} or equivalently by setting $p'(\psi)=0$ in the expression for MHS force balance \eqref{eq: complete surface sln with MHS}. Using that $B=1/\mu\rho$ is independent of $\alpha$, we integrate this to find
\begin{equation}
    \lambda = \frac{1}{B(l,\psi)} \int^l dl' \partial_\psi B(l',\psi)  + \frac{f(\psi)}{B(l,\psi)},
\end{equation}
where $f$ is the integration constant and is independent of $\alpha$ because we have $\lambda_\alpha = 0$, as in equation \eqref{eqs: complete surface solution 0}. We now consider the transformation from $(l,\alpha,\psi)$ coordinates to $(\phi,\alpha,\psi)$, where $\phi$ is the toroidal Boozer angle. We thus have
\begin{align}
    \left( \frac{\partial}{\partial \psi}\right)_l &= \left( \frac{\partial}{\partial \psi}\right)_\phi  + \left(\frac{\partial \phi}{\partial \psi}\right)_l \left( \frac{\partial}{\partial \phi}\right)_\psi\\
    &= \left( \frac{\partial}{\partial \psi}\right)_\phi  + \left( \frac{1}{G} \int dl' \partial_\psi B(l',\psi) \right) \left( \frac{\partial}{\partial \phi}\right)_\psi .
\end{align} 
The second equality follows from noting that in vacuum, the magnetic field can be written as $\boldsymbol{B}=G\bnabla \phi$, where $G$ is constant. In other words, $G\phi$ is the magnetic potential for a vacuum magnetic field. Rearranging this expression and using $\partial_\phi = (G/B) \partial_l$ gives
\begin{equation}\label{eq: partial psi l to phi}
    \left( \frac{\partial}{\partial \psi}\right)_\phi  = \left( \frac{\partial}{\partial \psi}\right)_l - \lambda \left( \frac{\partial}{\partial l}\right)_\psi, 
\end{equation}
where we have absorbed the constant $f(\psi)$ into the choice of origin between $l$ and $\phi$.

Now, we return to the fact that the current in the $\partial_\alpha\boldsymbol{r}$ direction vanishes: $(\lambda/\mu\rho)_l = (1/\mu\rho)_\psi$. We combine this with \eqref{eqs: complete surface solution a} to find
\begin{align}
    \frac{\kappa}{\rho} &= \left(\partial_\psi -  \lambda \partial_l \right) B\\
    &= \left( \frac{\partial B}{\partial \psi}\right)_\phi,\label{eq:kapparho flat mirror}
\end{align}
where we have used equation \eqref{eq: partial psi l to phi} in the second equality. For a foliation of Hasimoto surfaces, $\kappa/\rho=\gamma(\psi)$, so we find that
\begin{equation}\label{eq: field strength flat mirror}
    B(\phi,\psi) = B_0(\psi) + B_1(\phi), 
\end{equation}
where we use the fact that $B$ depends on only $\phi$ and $\psi$. This arises from equation \eqref{eqs: physical constraints c} and is equivalent to quasi-poloidal symmetry in the vacuum case considered here. We find that the magnetic field strength is separable into a radially varying part $B_0(\psi)$ and a surface-varying part $B_1(\phi)$, which is constant across flux surfaces. As an even stronger constraint on the field strength, this satisfies the flat mirror condition \eqref{eq: flat mirror condition}. Technically, for the benefits of a flat mirror to be gained, it is required that $\partial_\psi B_{00}>0$ \citep{velasco2023robust_flat_mirror}. However, this can often be achieved by adding a finite plasma pressure, due to the plasma behaving diamagnetically. We have indeed concluded that vacuum foliations of Hasimoto surfaces, which are a subset of QP magnetic fields, are flat mirrors. It is important to note that this property does not follow from QP generally, but just this particular class of QP solutions. The optimisation problem sketched in \S\ref{subsec: properties of surface equations} can be easily modified to look for Hasimoto surfaces, which then provides a 2D method for investigating flat mirror QP configurations.

In addition to the flat mirror result, we now derive a differential equation for the magnetic field strength in terms of the surface geometry quantities. In principle, this allows us to express quantities of practical interest, such as the mirror ratio, in terms of the surface geometry. We assume that we have a toroidal Hasimoto surface and that $(\rho/\kappa)_{l \psi}=0$, which is consistent with assuming $\rho \propto \kappa$ in the neighbourhood of the Hasimoto surface. We do not assume here that we have vacuum. We first note that equation \eqref{eqs: complete surface solution b} can be written as
\begin{equation}\label{eq: nu rho l deriv}
    \left(\frac{\nu}{\rho}\right)_l = -2 \frac{\mu \tau}{\rho}.
\end{equation}
Also, from equations \eqref{eq: rho_div_kappa_constraint}, \eqref{eqs: complete surface solution e}, \eqref{eqs: complete surface solution d}, we have
\begin{equation}\label{eq: nu rho alpha deriv}
    \left( \frac{\nu}{\rho}\right)_\alpha=\left( \partial_\psi - \lambda \partial_l\right) \ln\frac{\rho}{\kappa} + \left( \frac{4 \tau^2}{\kappa} + \kappa - \frac{\rho_{ll}}{\rho\kappa}\right) \frac{1}{\rho B} + \frac{1}{\kappa} \left( \frac{1}{\rho B}\right)_{ll},
\end{equation}
where we have used $B=1/\mu \rho$, which follows from the definition of the field-line-following coordinate system described in \S\ref{sec: moving frame representation}. We now equate the $\alpha$ derivative of equation \eqref{eq: nu rho l deriv} with the $l$ derivative of equation \eqref{eq: nu rho alpha deriv}---another compatibility condition. This gives
\begin{equation}
    - \left( \frac{2 \tau }{B \rho^2}\right)_\alpha = \left [\left( \frac{4 \tau^2}{\kappa} + \kappa - \frac{\rho_{ll}}{\rho\kappa}\right) \frac{1}{\rho B} + \frac{1}{\kappa} \left( \frac{1}{\rho B}\right)_{ll} \right]_l,
\end{equation}
where we have used the fact that $(\rho /\kappa)_l=0$ and $(\rho/\kappa)_{l \psi}=0$, consistent with field lines obeying the vortex filament equation on the surface and in the neighbourhood of the surface. We now integrate this equation around an integral curve of $\boldsymbol{u}=\partial_\alpha \boldsymbol{r}$, which is closed in toroidal geometry. Thus, as $2 \tau/B\rho^2$ is single-valued, the left-hand side vanishes and we have
\begin{equation}\label{eq: hasimoto surface field strength}
    \left(\frac{1}{B}\right)_{ll}\oint d\alpha \frac{1}{\rho \kappa} -2 \left(\frac{1}{B}\right)_l \oint d\alpha \frac{\rho_l}{\kappa \rho^2} + \frac{1}{B} \oint d\alpha \frac{1}{\rho \kappa}\left( 4\tau^2 + \kappa^2 - \frac{2\rho_{ll}}{\rho} + \frac{2 \rho_l^2}{\rho^2} \right)  = C, 
\end{equation}
where $C$ is some constant and we have used $\partial_\alpha B = 0$. The notable feature of this equation is that it establishes a direct relationship between the magnitude of the magnetic field and the geometry of the field lines, even though the exact nature of this relationship is obscure.

In this subsection, we have used $\kappa/\rho = \gamma(\psi)$ not only on the surface but also in its neighbourhood. In doing so, extra consistency conditions arise by combining equations \eqref{eqs: complete surface solution e} and \eqref{eqs: complete surface solution d}. We have not considered the consequences of enforcing this consistency. 

\section{Comparison with optimised numerical QP magnetic fields}\label{sec: numerical verification}

In this section, we present numerically generated vacuum QP magnetic fields, with the primary purpose of corroborating the surface equations. We caution the reader that we do not implement the proposed optimization scheme of \S\ref{subsec: properties of surface equations}, but optimize 3D fields for QP, from which we compute the surface quantities. We present two sets of toroidal magnetic field configurations, with specifications detailed in table \ref{tab:equilibria specs}. These configurations were chosen because they are sufficiently varied to show the surface equations in different contexts, they exhibit sufficiently accurate QP to show regions where the surface equations are true, and they also fairly show where breakdowns can, and often do, occur.

\begin{figure}
    \centering
    \includegraphics[width=\linewidth]{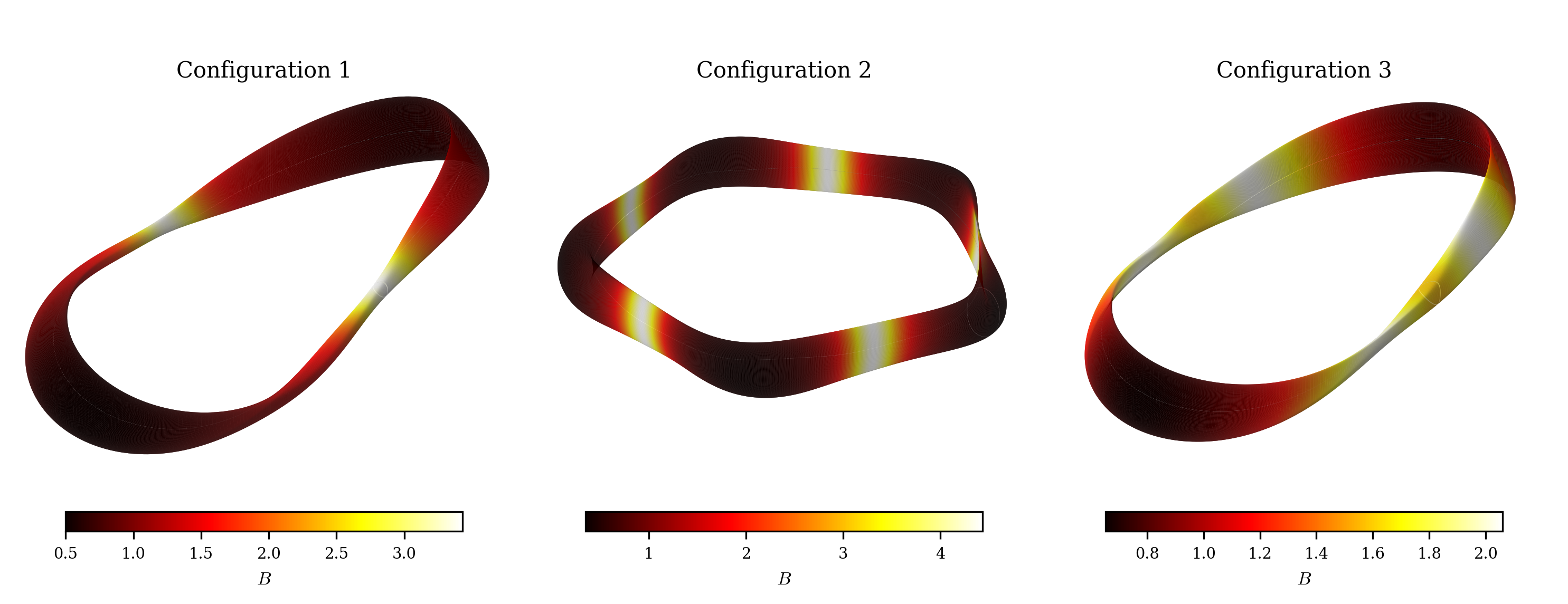}
    \caption{The boundary surface of the three configurations produced via {\tt DESC}.}
    \label{fig:3d_desc}
\end{figure}

The first three magnetic field configurations were produced and optimised for using the {\tt DESC} code \citep{Dudt2020}. Their boundary surfaces are displayed in figure \ref{fig:3d_desc}. By the nature of the method for calculating the equilibrium field, this equilibrium necessarily has nested flux surfaces. Each configuration was produced by optimising for QP on the boundary (not throughout the whole volume), alongside an aspect ratio and rotational transform target on the boundary, and a vacuum target throughout the volume. The configurations were optimized via the proximal least-squares method. Furthermore, all configurations are converged, satisfying functional tolerance and gradient tolerance criteria. The weights in the objective function were chosen as the best performing from a set of five randomly-generated sets. The initial condition for the optimisation came from the same class as that in section 3 of \cite{Goodman23}.

We also present two magnetic field configurations which were solved for by the {\tt SPEC} code \citep{Hudson2012}. They were optimised for using the adjoint methods developed in \cite{Nies_Paul_Hudson_Bhattacharjee_2022}. These configurations are exactly vacuum throughout the whole volume, which comes at the cost of not necessarily having nested flux surfaces. As with the other set, these configurations were generated by optimising for QP on the boundary, alongside an aspect ratio and rotational transform target. The initial condition for the optimisation was a rotating ellipse with the desired aspect ratio and rotational transform, as predicted by \cite{Mercier_1964}. In both sets of configurations, the optimisation took place in steps, with successive increases to the number of boundary modes \citep{landreman22}.

In the following subsections, we comment on the quality of quasi-poloidal symmetry (\S\ref{subsec: quality of QP}), we detail how the surface equations are tested (\S\ref{subsec: how the surface eqns are tested}), and then we explain deviations from the surface equations through the appearance of cusps (\S\ref{subsec: appearance of cusps}), non-zero parallel current (\S\ref{subsec: non-zero parallel current}), and loss of geodesicity (\S\ref{subsec: When are Field Lines Sufficiently Geodesic}). Finally, we note the preference for high mirror ratios, double wells, and narrow pinch points in \S\ref{subsec: high mirror ratios and narrow pinch points}.

\begin{table}
    \centering
    \begin{tabular}{C{2cm}C{2cm}C{2cm}C{2cm}C{2cm}C{2cm}}
         Configuration&  MHS Solver&  AR&  $\iota$&  $R$ & NFP\\ \midrule
         1&  {\tt DESC}& 14.12 & 0.30 & 0.74 & 2\\
         2&  {\tt DESC}&  11.36 & 0.11 & 0.85 & 5 \\
         3&  {\tt DESC}& 14.09 & 0.20 & 0.51 & 2\\
         4&  {\tt SPEC}& 6.02 & 0.27 & 0.97 & 5\\
         5&  {\tt SPEC}& 6.00 & 0.62 & 0.97 & 5 \\ 
         \bottomrule
    \end{tabular}
    \caption{We present specifications of the five numerical, QP equilibria, including the aspect ratio AR, the rotational transform $\iota$ on the boundary, the mirror ratio $R=(\text{max}(B)-\text{min}(B))/(\text{max}(B)+\text{min}(B))$ on the boundary, and the number of field periods, NFP.}
    \label{tab:equilibria specs}
\end{table}

\subsection{Quality of quasi-poloidal symmetry}\label{subsec: quality of QP}

We define the QP error on a surface of a configuration as 
\begin{equation}\label{eq: QP error def}
    \text{QP error}=\frac{\int_0^{2 \pi} d\vartheta \int_0^{2 \pi} d\varphi \sqrt{g}_\text{surf} \frac{1}{B^3} \left| \boldsymbol{B}\times \bnabla B \cdot \bnabla \psi \right|}{\int_0^{2 \pi} d\vartheta \int_0^{2 \pi} d\varphi \sqrt{g}_\text{surf}},
\end{equation}
where $2\pi\psi$ is the toroidal magnetic flux and $\int_0^{2 \pi} d\vartheta \int_0^{2 \pi} d\varphi \sqrt{g}_\text{surf} \left(\dots \right)$ is an integral over the surface, given that $(\vartheta,\varphi)$ are the computational poloidal and toroidal angle coordinates (not necessarily straight field-line coordinates) and $\sqrt{g}_\text{surf} = | \partial \boldsymbol{r}/\partial \vartheta \times \partial \boldsymbol{r}/\partial \varphi |$ is the surface Jacobian. This QP error is based on the commonly used `two-term' formulation of quasisymmetry (see, e.g., \citealt{Rodriguez_Paul_Bhattacharjee_2022}), but there is only one term because we aim for QP configurations which have zero enclosed toroidal current. The QP objective function used during the optimisations was slightly different to this QP error definition and is spelled out in appendix \ref{apdx: qp objective targets}.

\begin{figure}
    \centering
    \includegraphics[width=0.7\linewidth]{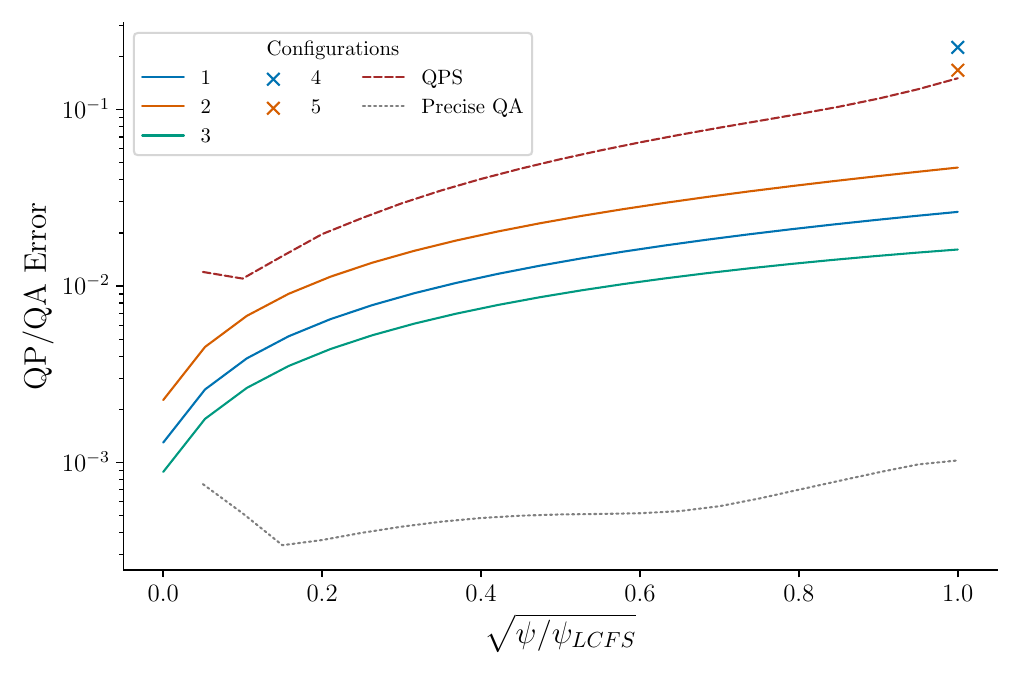}
    \caption{The QP error (defined in equation \eqref{eq: QP error def}) for six different magnetic field configurations as a function of the flux surface and the QA error (defined in equation \eqref{eq: QA error def}) for one additional configuration. Configurations 1-3 were produced using {\tt DESC} and configurations 4 and 5 were produced using {\tt SPEC} (QP error plotted only on boundary where a flux surface necessarily exists). `QPS' corresponds to the magnetic field design for the Quasi-Poloidal Stellarator \protect{\citep{NELSON2003205,Spong_2005}} and `precise QA' corresponds to the configuration found in \cite{landreman22}. Configurations 1-3 show an improvement over QPS. Despite QP being only targeted on the boundary, there is a lower level of QP towards the axis.}
    \label{fig: QP error}
\end{figure}

For the five configurations, we present the quality of quasi-poloidal symmetry as a function of the surface in figure \ref{fig: QP error}. We also make the comparison with the magnetic field design for the proposed Quasi-Poloidal Stellarator Experiment (QPS) \citep{NELSON2003205,Spong_2005} and the vacuum, precise QA configuration of \cite{landreman22}. For the latter, the error is appropriately adjusted to be 
\begin{equation}\label{eq: QA error def}
    \text{QA error}=\frac{\int_0^{2 \pi} d\vartheta \int_0^{2 \pi} d\varphi \sqrt{g}_\text{surf} \frac{1}{B^3} \left| \boldsymbol{B}\times \bnabla B \cdot \bnabla \psi + \frac{G}{\iota}\boldsymbol{B}\cdot\bnabla B\right|}{\int_0^{2 \pi} d\vartheta \int_0^{2 \pi} d\varphi \sqrt{g}_\text{surf}},
\end{equation}
where $2 \pi G$ is the external poloidal current. 

In figure \ref{fig: QP error}, we see that the level of QP error is significantly higher than that of the precise QA configuration, despite the {\tt DESC} configurations representing an improvement over QPS. We do not intend for this work to represent the forefront of possible QP solutions (see \citealt{haque2026} instead), as these solutions were generated for the purpose of corroborating the surface equations. Nonetheless, it seems that optimising for QP using standard methods is more difficult than for QA or QH. For solutions outside the region of validity of near-axis, this difficulty has not been explained. It might be due to a limitation of current optimization methods, in which case the scheme of \S\ref{subsec: properties of surface equations} could lead to improved QP solutions. It may also be due to fundamental difficulties with QP. Indeed, as we detail in \S\ref{subsec: appearance of cusps}, there is a fundamental tension between the third surface equation and toroidicity that can lead to cusps in the surface.

\subsection{How the surface equations are tested}\label{subsec: how the surface eqns are tested}

The first two surface equations, \eqref{eq: codazzi-mainardi a} and \eqref{eq: codazzi-mainardi b}, are purely geometric and are satisfied by any embedded surface foliated by geodesic field lines. Thus, to verify the first two surface equations, it suffices to ensure that the geodesic curvature is negligible on the boundary surface. As it turns out, this is a direct target for the optimiser as the geodesic curvature is proportional to the local QP error on a surface:
\begin{equation}\label{eq: two-term is geodesics}
    \frac{1}{B^3}  \boldsymbol{B}\times \bnabla B \cdot \bnabla \psi  = \rho \kappa_g,
\end{equation}
(cf., equation \eqref{eq: QP error def} and the optimiser objective functions presented in appendix \ref{apdx: qp objective targets}). That $\kappa_g \rightarrow 0$ leads to QP was explained in \S\ref{subsec: relation to QP}: a surface with geodesic field lines, zero net enclosed toroidal current, and zero normal current is necessarily QP. Despite being a direct target of the optimiser, deviations from zero geodesic curvature do exist and are significant, as demonstrated in \S\ref{subsec: appearance of cusps}. 

The third surface equation, \eqref{eq: rho_div_kappa_constraint}, is assessed directly. For this, we need to interpret $\kappa$, $\rho$, and $\partial_\alpha$ for our numerical equilibria. We take $\kappa$ to be the normal curvature, $\kappa_n$. For calculating $\rho$, we use the transformation to Boozer coordinates
\begin{subequations}\label{eqs: numerical transform to boozer}
    \begin{align}
        \theta &= \vartheta + \tilde{\lambda} + \iota \omega,\\
        \phi &= \varphi + \omega,
    \end{align}
\end{subequations}
where $\tilde{\lambda}(\vartheta,\varphi)$ gives the transformation to straight field-line coordinates and $\omega(\vartheta,\varphi)$ gives the transformation from those coordinates to Boozer coordinates. Calculating $\tilde{\lambda}$ and $\omega$ from numerical equilibria is standard practice when provided with flux surfaces (since $\boldsymbol{\bnabla}\psi$ can be straight-forwardly defined). For the {\tt SPEC} equilibria, there is no obvious globally-defined $\bnabla \psi$ since there are no nested flux surfaces ---- see \cite{Nies_Paul_Hudson_Bhattacharjee_2022} for how  $\tilde{\lambda}$ and $\omega$ are calculated in this case. In vacuum, the magnetic field can be written as $\boldsymbol{B}=G\bnabla(\omega + \varphi)$, where $G$ is constant. Given the surface Jacobian $\sqrt{g}_\text{surf} = | \partial \boldsymbol{r}/\partial \vartheta \times \partial \boldsymbol{r}/\partial \varphi |$, we then take $\rho$ to be
\begin{equation}\label{eq: numerical calculation of rho}
    \rho =  \frac{\sqrt{g}_\text{surf} B}{G\left| \left(1+\omega_\varphi\right) \left(1+ \tilde{\lambda}_\vartheta\right) - \omega_\vartheta \left(\tilde{\lambda}_\varphi - \iota \right)  \right|},
\end{equation}
where $\iota$ is the rotational transform. This agrees with the theoretical definition of $\rho$ when $\kappa_g=0$, as shown in appendix \ref{apdx: numerical calc of rho}.

\begin{figure}
    \centering
    \includegraphics[width=1\linewidth]{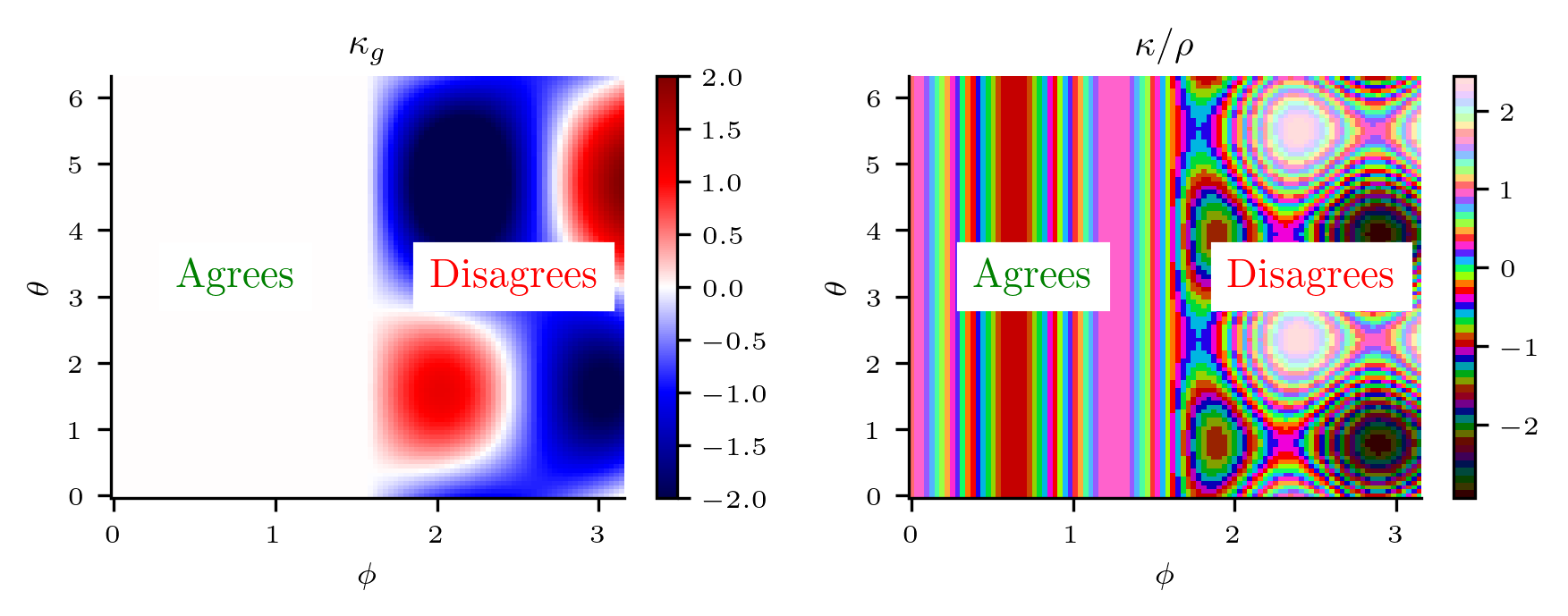}
    \caption{A schematic example of what it looks like for the surface equations to be verified or falsified on a flux surface. For $\phi < \pi/2$, there is agreement with the surface equations, since the geodesic curvature is zero and $\kappa/\rho$ only depends on $\phi$. For $\phi > \pi/2$, there is disagreement as the geodesic curvature is non-zero and $\kappa/\rho$ does not have vertical contours.}
    \label{fig: surface eqns test schematic}
\end{figure}

Finally, for exact QP we have
\begin{equation}
    \left( \frac{\partial}{\partial \theta} \right)_\phi = 0 \quad \iff\quad \left(\frac{\partial}{\partial \alpha}\right)_l=0,
\end{equation}
so the third surface equation is verified if and only if $\kappa/\rho$ depends only on $\phi$. In figure \ref{fig: surface eqns test schematic}, we illustrate what the plots of $\kappa_g$ and $\kappa/\rho$ on the boundary should look like for the surface equations to be verified.

\subsection{The appearance of cusps}\label{subsec: appearance of cusps}

\begin{figure}
    \centering
    \includegraphics[width=1\linewidth]{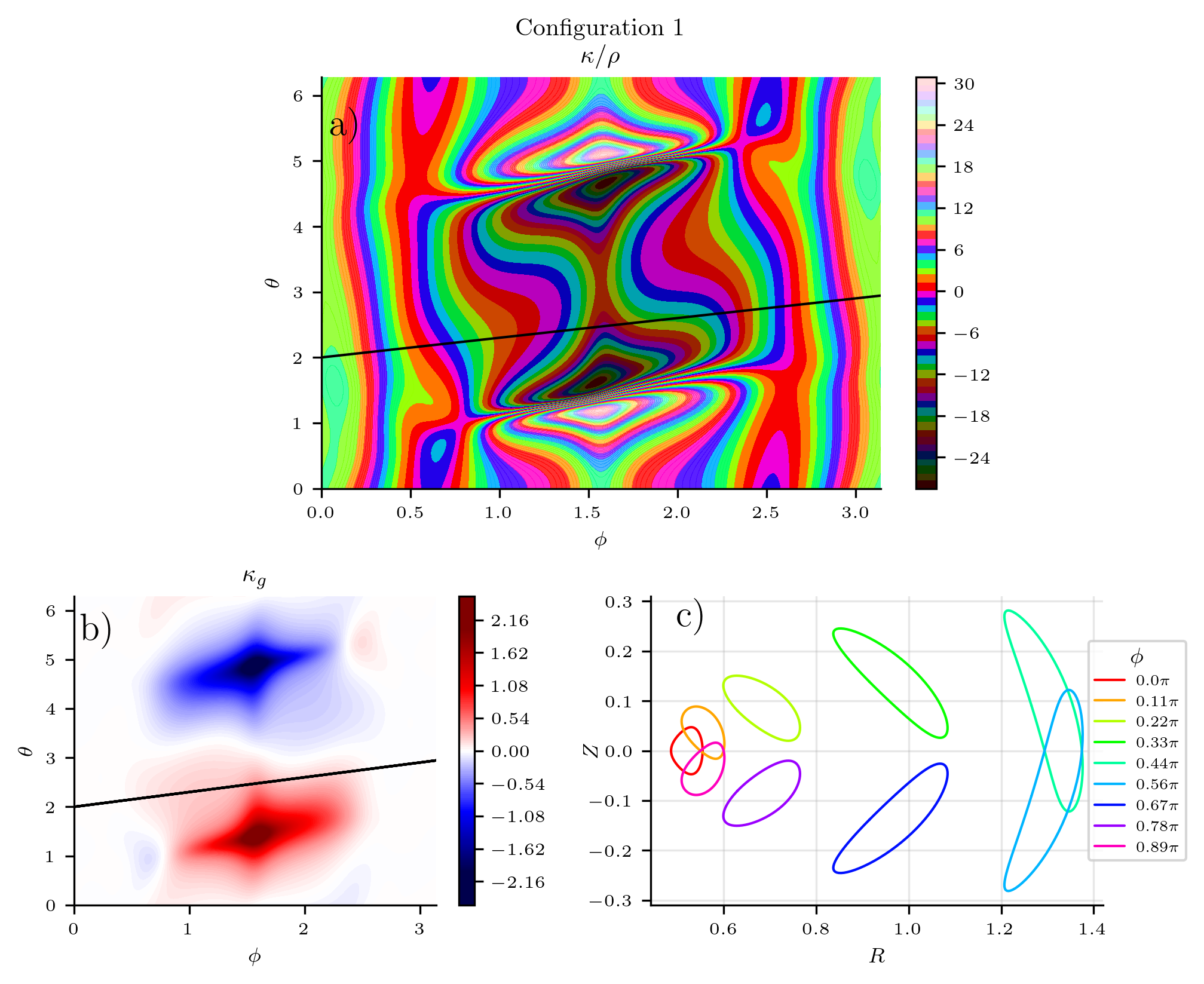}
    \caption{We plot quantities for the first {\tt DESC} configuration: a) $\kappa/\rho$ and b) the geodesic curvature $\kappa_g$, both on the boundary surface as a function of Boozer angles, and c) toroidal cross-sections of the boundary flux surface. Lengths are measured with respect to the average major radius. The black line in a) and b) is a field line. We see agreement with the third surface equation, except the regions with large geodesic curvature (i.e., where QP-breaking is significant).}
    \label{fig: config 1 cusps}
\end{figure}

In this subsection, we consider the first {\tt DESC} magnetic field configuration. For this configuration, the vacuum target was essentially achieved. In figure \ref{fig: config 1 cusps}a), we see that $\kappa/\rho$ varies primarily horizontally in agreement with the third surface equation, except for regions of sharp vertical change around $\theta \approx 1$ and $\theta \approx 4.5$ for $\phi \in (0.75,2.5)$. This does not falsify the third surface equation because these are precisely the regions where the geodesic curvature deviates significantly from zero, as can be seen from figure \ref{fig: config 1 cusps}b). After all, zero geodesic curvature was an assumption behind the derivation of the surface equations.

In fact, we can understand these deviations from the surface equations through the surface equations themselves. The curvature $\kappa$ is negative whenever the field line principal normal is opposed to the surface normal. Based on intuition from tokamaks, the inboard and outboard sides of the torus typically have opposite signs of $\kappa$. However, the third surface equation tells us that $\kappa/\rho$ should be constant on poloidal loops, so $\kappa$ should not be able to change sign, since $\rho>0$. The optimiser rectifies this difficulty by isolating the regions where $\kappa$ changes sign to thin cusps, with $\kappa/\rho$ independent of $\theta$ only away from these cusps. The cusps are visible in figure \ref{fig: config 1 cusps}c) at $\phi \approx 0.5 \pi$ as the tips on the top and bottom of the cross-section. These cusps are also visible in figure \ref{fig:3d_desc} where the surface `curves around'.  

We note that the cusps also appear to follow the field lines. This can be understood through MHS force balance. As the field is vacuum, it is necessarily force-free, so the tension force from the field lines can only be balanced by magnetic pressure. The magnetic pressure can only grow so large given the limited field strength variation, so the magnetic tension force has to be minimised by not having field lines go across the cusp.

\subsection{The effect of parallel current}\label{subsec: non-zero parallel current}

\begin{figure}
    \centering
    \includegraphics[width=1\linewidth]{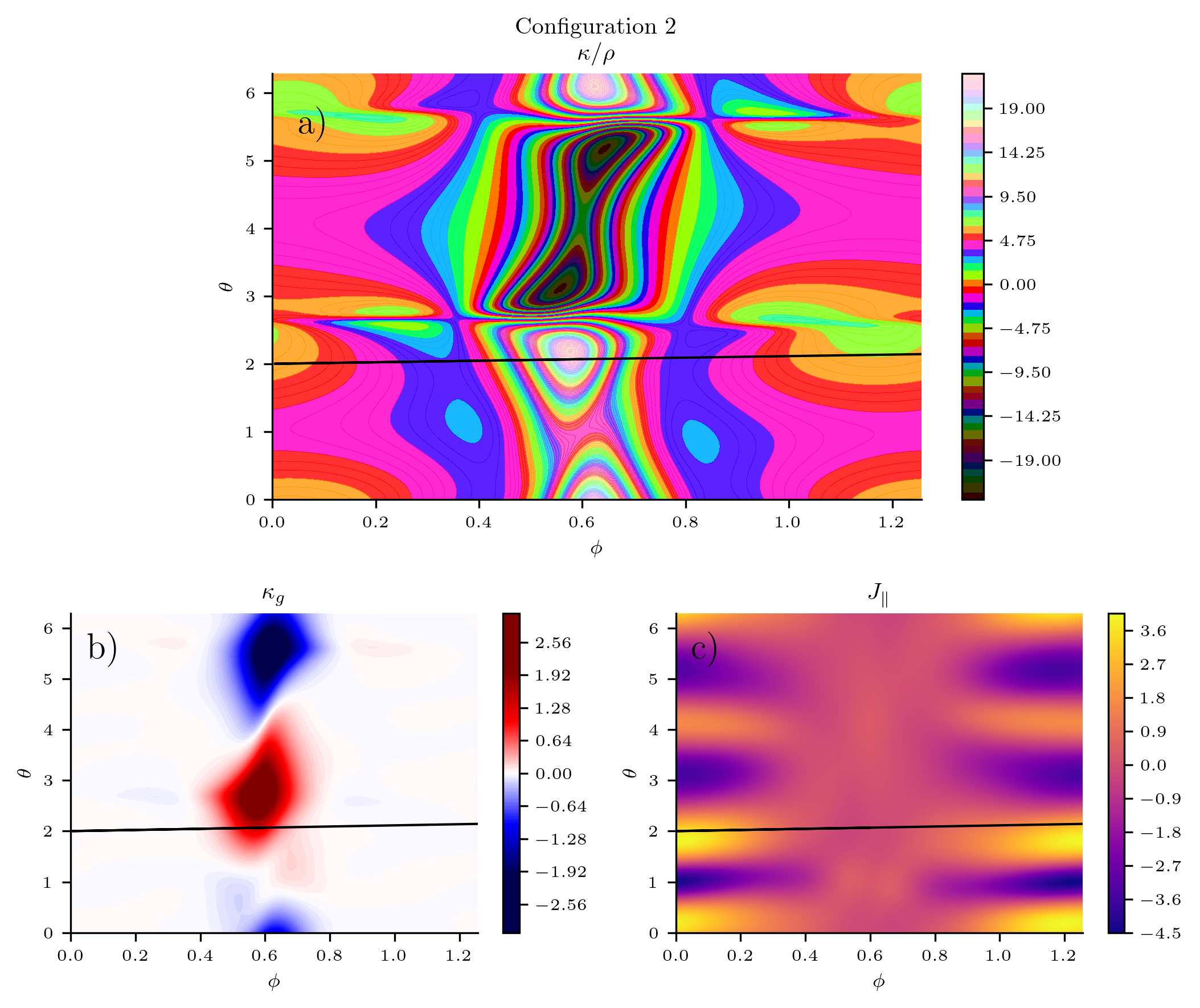}
    \caption{For the second {\tt DESC} configuration: a) $\kappa/\rho$, b) the geodesic curvature $\kappa_g$,  and c) the parallel current $J_\parallel$. All quantities are plotted on the boundary surface as a function of Boozer angles. Lengths are measured with respect to the average major radius and the current is normalised by the mean magnetic field on the boundary. There are two related reasons for deviations from the third surface equation: areas of high geodesic curvature and large parallel current.}
    \label{fig: config 2 current}
\end{figure}

We now consider the second {\tt DESC} magnetic field configuration. In figure \ref{fig: config 2 current}a), we observe noticeable departure from the third surface equation, since the contours are largely not close to vertical. There are two related reasons for this.

Firstly, consider the region $\phi \in (0.4,0.8)$. This region of the surface is similar to the configuration considered in the last subsection: there are approximately straight contours separated by `cusps', which is where the geodesic curvature is largest (see figure \ref{fig: config 2 current}b)). The geodesic curvature deviates somewhat from zero even away from these cusps, which explains why the vertical contours are slightly curved and angled. As in the last subsection, we can explain the cusps by the fact that $\kappa$ typically has to change sign from the inboard to the outboard side, in contradiction with the third surface equation.

Now, consider the remaining portion of the surface: $\phi \in (0,0.4)$ and $\phi \in (0.8,2\pi/5)$. The contours are not vertical here at all, despite the geodesic curvature being small. Here, it is another assumption of the surface equations which breaks down: the parallel current is non-zero. Zero current density was a target for the optimiser but was not sufficiently attained everywhere in this configuration.

Even without considering the vacuum target, having non-zero parallel current is somewhat unexpected. This is because of the theoretical result presented at the end of appendix \ref{apdx: Boozer coords}: assuming QP, force-balance, and zero net toroidal current, the parallel current should vanish pointwise. The magnetic field configuration presented in figure \ref{fig: config 2 current} does indeed have zero net toroidal current and is force-balanced (as current perpendicular to the field lines is indeed negligible). Thus, it is deviations from QP (i.e. deviations from zero geodesic curvature) which leads to non-zero parallel currents. Curiously, the region where geodesic curvature is significant ($\phi \in (0.4,0.8)$) is different to the region where parallel currents are significant: hence, this is a non-local effect.

\subsection{When are field lines sufficiently geodesic?}\label{subsec: When are Field Lines Sufficiently Geodesic}

\begin{figure}
    \centering
    \includegraphics[width=1\linewidth]{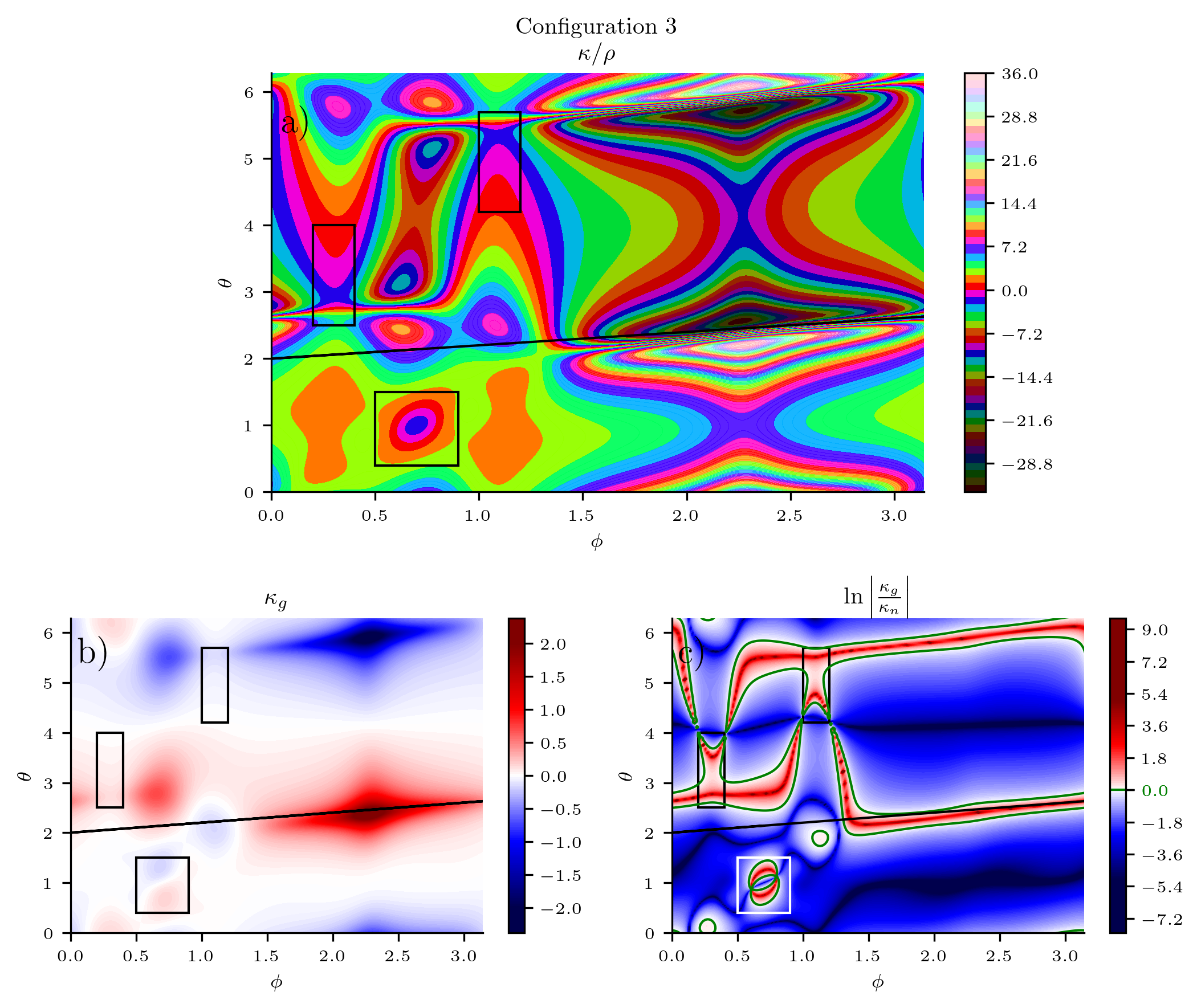}
    \caption{For the boundary surface of the third {\tt DESC} configuration: a) $\kappa/\rho$, b) the geodesic curvature $\kappa_g$,  and c) the logarithm of the geodesic curvature divided by normal curvature $\kappa_n$. For the third surface equation to be true, the geodesic curvature should be small compared to the normal curvature, which is not the case within the overlaid black/white rectangles.}
    \label{fig: config 3 geodesicity}
\end{figure}

In this subsection, we use the third {\tt DESC} magnetic field configuration to point out that the third surface equation is satisfied (locally) only if the field lines are sufficiently geodesic, in the sense that the geodesic curvature should be much smaller than the normal curvature.

We consider two different parts of the boundary surface of this configuration. The region $\phi\in(1.3,\pi)$ is similar to the central region of the first configuration considered in \S\ref{subsec: appearance of cusps}: the third surface equation is violated on field-line-following cusps (see figure \ref{fig: config 3 geodesicity}a)), which is where the geodesic curvature (and thus QP error) is large, as per figure \ref{fig: config 3 geodesicity}b). Away from these cusps, the contours of $\kappa/\rho$ are nearly vertical.

We now consider the region $\phi \in (0,1.3)$, where the magnetic field strength is largest. Apart from distinct blobs at $(\phi,\theta)\approx(0.7,3)$ and at $(\phi,\theta)\approx(0.7,5.5)$, the geodesic curvature is relatively small in an absolute sense, as per figure \ref{fig: config 3 geodesicity}b). These blobs can also be thought of as cusps, and also represent areas where the third surface equation does not hold (contours of $\kappa/\rho$ are not vertical). However, even away from these blobs, the contours are not close to vertical, particularly in the regions outlined by the black rectangles in figure \ref{fig: config 3 geodesicity}a). As seen in figure \ref{fig: config 3 geodesicity}b), the geodesic curvature is small in these regions, in an absolute sense. However, as seen from figure \ref{fig: config 3 geodesicity}c), the geodesic curvature is much larger than the normal curvature in these regions, so the field lines cannot be considered geodesics in these regions. Given that the geodesicity of field lines was an assumption of the surface equations, deviation from the third surface equation is to be expected in these regions.

Given that the normal curvature does not directly enter into the definition of QP error (cf., equations \eqref{eq: QP error def} and \eqref{eq: two-term is geodesics}), this observation suggests that it is theoretically possible to have a surface of high (but imperfect) apparent QP quality without having that surface conform to the surface equations, as long as the normal curvature is sufficiently small. This would be most readily achieved for a high aspect ratio stellarator, where it is conceivable for normal curvature to remain small everywhere. However, this would constitute a Pyrrhic victory. As discussed in \S\ref{sec: introduction}, many practical advantages stem from the radial drift of the guiding centres being much smaller than the poloidal drift. If geodesic curvature is larger than normal curvature, the radial drift is larger than the poloidal drift, so one loses these practical advantages.

\subsection{High mirror ratios and narrow pinch points}\label{subsec: high mirror ratios and narrow pinch points}

\begin{figure}
    \centering
    \includegraphics[width=1\linewidth]{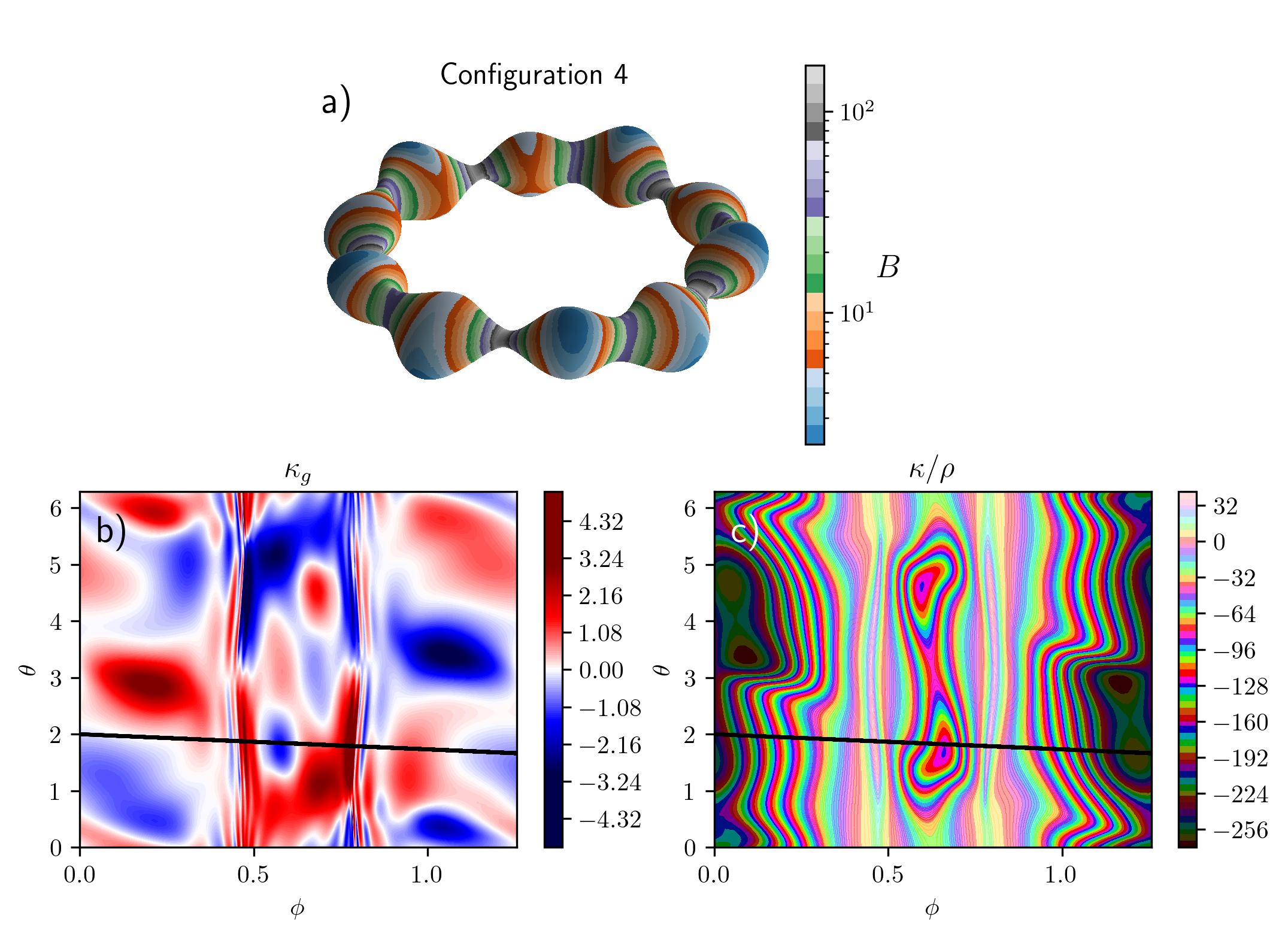}
    \caption{For the first {\tt SPEC} configuration, we plot a) the boundary surface in real space, b) $\kappa_g$ on the boundary, and c) $\kappa/\rho$ on the boundary. Lengths are measured in terms of the average major radius and the black line on b) and c) is a field line. We observe a very large mirror ratio, narrow pinch points, and double wells.}
    \label{fig: spec config 4}
\end{figure}

\begin{figure}
    \centering
    \includegraphics[width=1\linewidth]{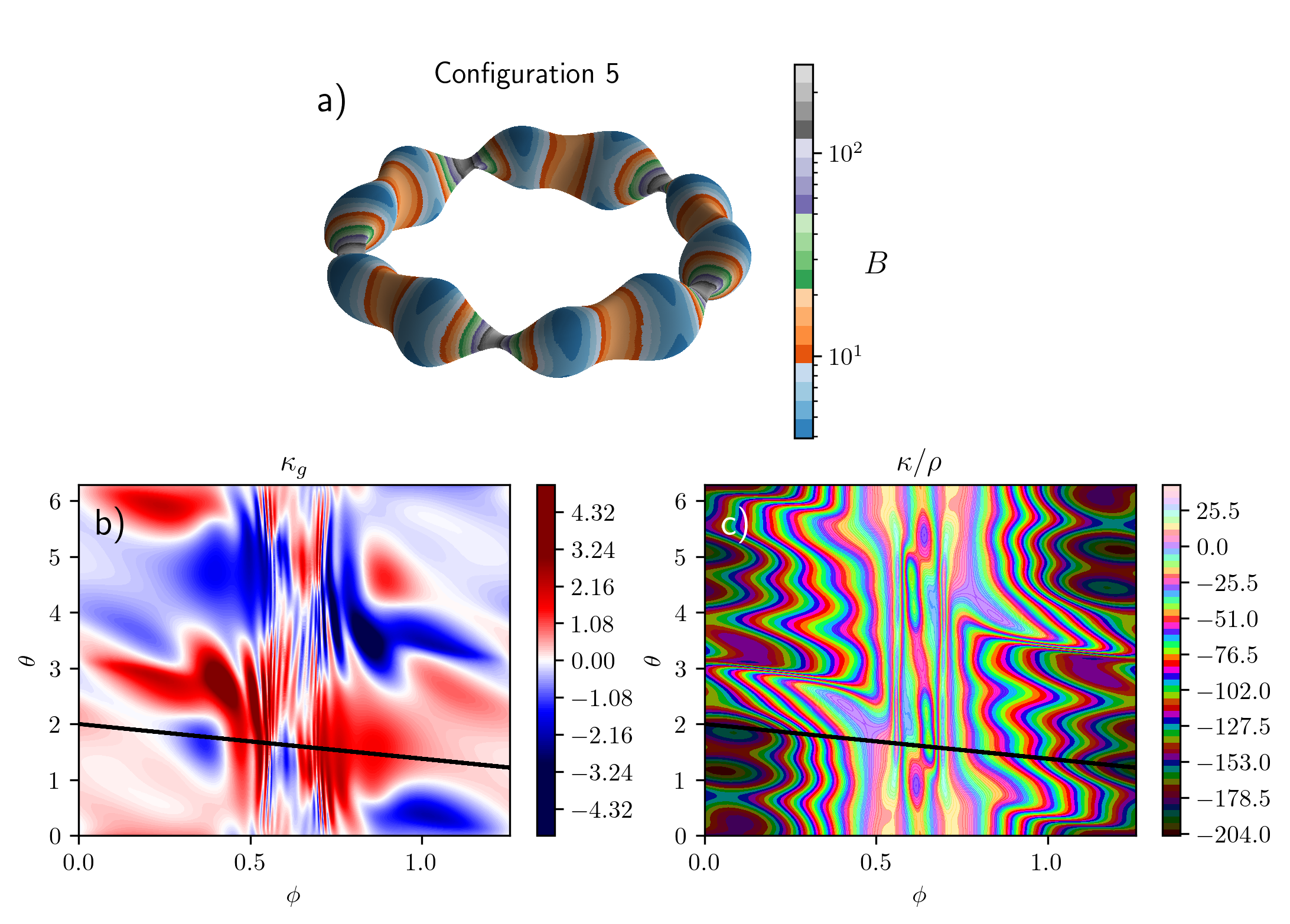}
    \caption{For the second {\tt SPEC} configuration, we plot a) the boundary surface in real space, b) $\kappa_g$ on the boundary, and c) $\kappa/\rho$ on the boundary. Lengths are measured in terms of the average major radius and the black line on b) and c) is a field line. We observe a very large mirror ratio, narrow pinch points, and double wells.}
    \label{fig: spec config 5}
\end{figure}

In this subsection, we focus on the two {\tt SPEC} equilibria. By construction, these are vacuum equilibria. Thus, we need not worry about non-zero currents undermining the surface equations, as was the case in \S\ref{subsec: non-zero parallel current}. This comes at the cost of not necessarily having nested flux surfaces, which permits pathologically narrow pinch points in the boundary surface, seen in figures \ref{fig: spec config 4}a) and \ref{fig: spec config 5}a). 

In figures \ref{fig: spec config 4} and \ref{fig: spec config 5}, we plot $\kappa/\rho$ and $\kappa_g$ on the boundary surface. Due to the regions of large geodesic curvature, there are deviations from the third surface equation as the contours of $\kappa/\rho$ are not vertical. There are vertical contours in select regions, particularly in figure \ref{fig: spec config 4}c). 

For completeness, we note an additional potential reason for deviation from the third surface equation. In deriving the surface equations, we assumed not only that the field lines are geodesics on the flux surface, but also that field lines are geodesics in the neighbourhood of the surface (see equation \eqref{eqs: physical constraints b} and surrounding discussion). In other words, we assumed QP in the neighbourhood of the surface. Thus, in these {\tt SPEC} configurations, non-geodesic field lines adjacent to the flux surface can contribute to deviations of the third surface equation. This differs from the {\tt DESC} configurations, where imposing nested flux surfaces naturally leads to a high level of QP on neighbouring surfaces.

One can observe in figures \ref{fig: spec config 4}a) and \ref{fig: spec config 5}a) that the {\tt SPEC} configurations resemble a system of `linked magnetic mirrors'. This observation is consistent with the surface equations, since magnetic mirrors are solutions to the surface equations with poloidally-closing orthogonals to the field lines.

We also observe narrow pinch points and extremely high mirror ratios in the {\tt SPEC} configurations, as seen in figures \ref{fig: spec config 4}a) and \ref{fig: spec config 5}a). Even with a prescribed number of field periods, the optimiser tended to `double' the field periods in these configurations. In other words, the optimisation would occasionally produce a `double well', with two minima of the magnetic field strength within a single field period.

Based on the similarity with magnetic mirrors, we can explain the appearance of narrow pinch points, high mirror ratios, and double wells through the surface equations. As it turns out, magnetic mirrors with narrow pinch points can more readily bend around (as is necessary to form a toroidal surface) and generate a non-zero rotational transform. This claim is justified in appendix \ref{apdx: perturbations to surfs of rev}, where we perturb surfaces of revolution through the surface equations, form an eigenvalue problem (given periodic boundary conditions), and show that the density of eigenvalues increases given a narrow pinch point.

In addition to this mathematical argument, we can understand physically why narrow pinch points assist in having non-zero rotational transform. The field lines are geodesics, which are paths of inertial motion on the surface. Akin to a ballerina pulling their arms in to spin faster, a narrower pinch point will assist the geodesic field lines `spinning around' the surface, generating a rotational transform. To consider an explicit example, consider a surface of revolution $\boldsymbol{r}(\phi,x)=(x,y(x)\text{cos}(\phi),y(x)\text{sin}(\phi)$). The geodesics are given by the solutions to 
    \begin{equation}
        1=(1+y'(x))^2\dot{x}^2+y(x)^2\dot{\phi}^2, \quad J=y^2\dot{\phi}.
    \end{equation}
for some constant J, which is the angular momentum here. One can see that for smaller $y$ at constant $J$, $\dot{\phi}$ has to be larger. This physical interpretation does not extend to explaining why narrower pinch points also assist in the surface being able to `bend around'. 

Finally, there are other possible explanations for the appearance of large mirror ratios and narrow pinch points in optimized QP stellarators. In the near-axis approximation, the curvature of the magnetic axis is proportional to the level of QP-breaking at first-order from the axis, as per equation \eqref{eq: near-axis curv leads to badQP}. The optimiser, in its pursuit to satisfy the QP metric, may swamp this necessary poloidal variation of the magnetic field with a large toroidal variation, leading to high mirror ratios. In addition, the near-axis expansion also indicates a preference for highly elongated surfaces, which also corresponds to high mirror ratios and narrow pinch points. High mirror ratios are physically undesirable: for instance, \cite{haque2026} finds that higher mirror ratios lead to increased fast ion losses, even with an apparently improved QP quality. Furthermore, \cite{Dommaschk_1994} provides a theoretical lower bound for the mirror ratio in QP stellarators. Thus, it is practically important to assess whether it is possible to achieve a high quality of QP without high mirror ratios. Appendix \ref{apdx: perturbations to surfs of rev} illustrates that it is consistent with the surface equations to have solutions with high mirror ratios when QP surfaces are close to surfaces of revolution, but it does not rule out the possibility of low-mirror-ratio QP solutions, which may be possible when the near-axis expansion breaks down.

To summarise this section, we have found that the surface equations are approximately obeyed where the local QP error, local parallel currents, and geodesic curvature are sufficiently small. It is important to note that these numerical solutions deviate significantly from ideal vacuum QP. These deviations often occur in local patches. These local patches serve as a reference case for where the surface equations need not be true and indeed are violated. The local QP error is proportional to the geodesic curvature, and we found that for the surface equations to be approximately satisfied, the geodesic curvature must be much smaller than the normal curvature. This is a physically motivated condition, since the ratio between geodesic curvature and normal curvature corresponds to the ratio between radial curvature drift and poloidal curvature drift. We observe a tendency for cusps to form on the surface, corresponding to areas of high geodesic curvature and deviations from the third surface equation. These cusps appear because of the optimiser's predicament: the third surface equation implies that $\kappa$ has a constant sign along the poloidal loops, but typical toroidal surfaces have $\kappa$ switching sign between the inboard and outboard sides. This represents a fundamental difficulty in achieving exact QP that applies outside the regime of validity of the near-axis framework. Finally, for QP surfaces that resemble linked magnetic mirrors, we found that narrow pinch points and high mirror ratios were consistent with the surface equations. It remains as important future work to verify whether high mirror ratios and narrow pinch points are essential to QP surfaces.

\section{Conclusions}
The main result of this paper is the surface equations \eqref{eqs: surface equations}, which represent a striking reduction of the problem of finding QP flux surfaces, as well as desirable asymmetric magnetic-mirror surfaces. Rather than solving the full 3D problem of finding a surface as well as neighbouring field lines, it suffices to solve the 2D problem of the surface equations. This comes with the limitation that nested, QP flux surfaces are not guaranteed. The surface equations were compared with numerical equilibria in \S\ref{sec: numerical verification}, where deviations from the surface equations correlated with local QP error (i.e., non-zero geodesic curvature) or significant parallel currents. These numerical equilibria were generated through optimising for QP and did not involve directly solving the surface equations.  The surface equations also provide a theoretical framework for understanding QP solutions in lieu of a near-axis expansion. For instance, we qualitatively explained why numerically optimised QP stellarators tend to have cusps: the optimiser attempts to reconcile the requirement that $\kappa/\rho$ remain constant on poloidal loops with the fact that $\kappa$ generally switches sign between the inboard and outboard sides of a toroidal surface. In addition, we identified two basic classes of solutions to the surface equations: generalised helicoids, which include magnetic mirrors but no toroidal solutions, and Hasimoto surfaces, which are surfaces traced out by a vortex filament. We observed that some numerical QP equilibria resemble linked magnetic mirrors and used this to explain the appearance of narrow pinch points and high mirror ratios. This explanation shows consistency with the surface equations, but we do not rule out the possibility that low-mirror-ratio QP solutions are possible. Such solutions would likely have to be low aspect ratio, since the near-axis expansion points towards high mirror ratios for QP. For toroidal Hasimoto surfaces, we developed a theory for the field strength, demonstrating that such surfaces in vacuum are flat mirrors; such configurations are of particular interest because they exhibit robust confinement of fast ions and small radial transport of energy. This analysis is limited, however, since toroidal Hasimoto surfaces (other than the \cite{palumbo1968_Closed_MHD} configuration) might not exist.

We outline directions for future work. The simplified, efficient optimisation problem, sketched out in \S\ref{subsec: properties of surface equations}, remains to be numerically implemented. This would allow for the production of a database of QP flux surfaces from which trends could be analysed. For instance, one could study whether high mirror ratios are intrinsic to QP, which could represent a possible unfortunate trade-off to designing a practical QP stellarator. It also remains to show whether exact QP flux surfaces are even possible, as no analytic solution has been presented in this paper. The surface equations provide a route to analytically investigating this question. That said, even if exact QP flux surfaces are impossible, the surface equations could still be of use in designing approximate QP surfaces, or QI surfaces with local regions of low geodesic curvature. The numerical solutions in this paper were primarily used for verifying the surface equations: other work \citep{haque2026} investigates numerically achieving as precise QP as possible and seeks to verify some of the associated physical benefits in these configurations, such as improved fast ion confinement. Further investigation is needed to verify some other attractive features of QP, such as the possible turbulence suppression from poloidal shearing flows. Finally, one could use the surface equations in designing a magnetic mirror that is MHD stable and has good radial, collisionless particle confinement. This would be particularly helpful for designs where the paraxial approximation (i.e. near-axis for magnetic mirrors; \citealt{Newcomb_1981}) is no longer applicable. However, it is cautioned that the considerations of this paper do not apply when there is current normal to the flux surfaces, which is possible in equilibria which possess significant pressure anisotropy.

\section*{Funding}
This research was supported by a grant from the Simons Foundation/SFARI (560651, AB) and the Department of Energy Award No. DE-SC0024548 (until March 31, 2025).

\section*{Declaration of interests}
The authors report no conflict of interest.

\appendix

\section{Impossibility of exact QP near the magnetic axis}\label{apdx: impossibility of QP near axis}
We reproduce the proof that quasi-poloidal symmetry is impossible near the magnetic axis \citep{landreman2011, helander_theory_2014}.

The curvature vector $\boldsymbol{\kappa} = \kappa \boldsymbol{\hat{n}}$ can be written as
\begin{align}
    \boldsymbol{\kappa} &= -\boldsymbol{\hat{t}}\times \left( \bnabla \times \boldsymbol{\hat{t}}\right)\\
    &=-\boldsymbol{\hat{t}}\times \left( \bnabla \frac{1}{B} \times  \boldsymbol{\hat{t}}\right) - \frac{\boldsymbol{\hat{t}}}{B}\times \boldsymbol{J}\\
    &=\frac{\bnabla p}{B^2}+\frac{\bnabla_\perp B}{B},
\end{align}
where we have used MHS force balance \eqref{eq: MHS force balance} and $\bnabla_\perp = \bnabla - \boldsymbol{\hat{t}} \boldsymbol{\hat{t}}\cdot \bnabla$. Close to the axis, $\bnabla \psi \rightarrow 0$. Thus, as $\bnabla p \propto \bnabla \psi$, we also have $\bnabla p \rightarrow 0$. In \cite{Boozer1981} coordinates $(\theta, \phi)$, the magnetic field can be written as $\boldsymbol{B} = G(\psi)\bnabla\phi + I(\psi) \bnabla \theta + \beta(\psi,\theta, \phi) \bnabla \psi$, where $2\pi \psi$ is the toroidal magnetic flux and $2\pi I(\psi)$ is the net toroidal current enclosed by the flux surface. Assuming the current parallel to the magnetic axis is not singular, we take $I \rightarrow 0$ as $\psi \rightarrow 0$, where $\psi=0$ on the axis. We also assume $\beta$ stays bounded for $\psi \rightarrow 0$. Thus, close to the axis, $\boldsymbol{B} = G(0) \bnabla \phi$. So, we have
\begin{equation}\label{eq: near-axis curv leads to badQP}
    \left|\boldsymbol{\kappa} \right| \propto \frac{\partial B}{\partial \theta},
\end{equation}
since $\partial_\theta \boldsymbol{r} \cdot \bnabla \phi = 0$. To form a closed loop, the axis has to curve somewhere, so at some point along the axis, $|\boldsymbol{\kappa}|\neq 0$. Thus, somewhere $\partial_\theta B \neq 0$, contradicting quasi-poloidal symmetry, which requires $B = B(\psi,\phi)$ by definition. There is a stronger version of this argument that shows QP is not possible close to the axis without invoking MHS force balance \citep{Plunk_Landreman_Helander_2019,Rodriguez2022}.

\textit{Prima facie}, this proof seems to dismiss quasi-poloidal symmetry as a possibility. However, a few provisos are in order. Firstly, this proof does not imply that QP cannot be achieved away from the axis, where having good QP is arguably more important. That said, some of these difficulties may also present away from the axis (see, e.g., \S\ref{subsec: appearance of cusps}). Notably, in a small aspect ratio stellarator, more of the stellarator volume will be away from the near-axis limit. Secondly, in the opposite limit of large aspect ratio, the curvature of the axis $|\boldsymbol{\kappa}|$ is small, so the level of QP-breaking can also be made small, as seen in equation \eqref{eq: near-axis curv leads to badQP}. Thus, this proof has limited consequences for both small aspect ratio and large aspect ratio stellarators. 

For completeness, we also list two technical ways to circumvent the proof. Firstly, one could have a singular current on the axis so that $I(\psi=0)\neq 0$. Secondly, the magnetic field on the axis can be made to vanish. These are not desirable possibilities for magnetic confinement.

\section{Equivalence of moving frame and Riemann curvature conditions}\label{apdx: riemann curvature}

Though we introduced $\boldsymbol{K},\boldsymbol{\chi}$, and $\boldsymbol{\sigma}$ in \S\ref{sec: moving frame representation}, physically we have just imposed the flatness of Euclidean space on the metric components, given by $\rho,\eta,\lambda,\mu,$ and $\nu$. The Riemannian curvature tensor is defined as
\begin{equation}\label{eq: def of riemann curvature tensor}
    \mathcal{R}(\boldsymbol{X},\boldsymbol{Y})\boldsymbol{Z} = \nabla_{\boldsymbol{X}} \nabla_{\boldsymbol{Y}} \boldsymbol{Z} - \nabla_{\boldsymbol{Y}} \nabla_{\boldsymbol{X}} \boldsymbol{Z} - \nabla_{[\boldsymbol{X},\boldsymbol{Y}]} \boldsymbol{Z},
\end{equation}
where $\nabla$ denotes the covariant derivative, $[\boldsymbol{X} ,\boldsymbol{Y}]= (X^j\partial_j Y^i - Y^j\partial_j X^i) \partial_i$ is the Lie bracket, and $\boldsymbol{X},\boldsymbol{Y},\boldsymbol{Z}$ are arbitrary vector fields \citep{riemann1953collected}. As an example, we take $\boldsymbol{X}=\partial\boldsymbol{r}/\partial l$, $\boldsymbol{Y}=\partial\boldsymbol{r}/\partial \alpha$, and $\boldsymbol{Z}=\boldsymbol{e}_i$, where we have defined $(\boldsymbol{e}_1,\boldsymbol{e}_2,\boldsymbol{e}_3)=(\boldsymbol{\hat{t}},\boldsymbol{\hat{n}},\boldsymbol{\hat{b}})$. Then, using the fact that the Lie bracket of coordinate fields vanishes, we have
\begin{equation}
        \mathcal{R} \left( \frac{\partial\boldsymbol{r}}{\partial l},\frac{\partial\boldsymbol{r}}{\partial \alpha} \right) \boldsymbol{e}_i = \left( \frac{\partial \chi_{ij}}{\partial l}  - \frac{\partial K_{ij}}{\partial \alpha} + \chi_{ik}K_{kj} - K_{ik}\chi_{kj} \right) \boldsymbol{e}_j,
\end{equation}
where we have used the skew-symmetric matrix versions of $\boldsymbol{K}$ and $\boldsymbol{\chi}$, i.e., $K_{ij} = \epsilon_{ijk} K_k$. This can be repeated for other combinations of the coordinate fields. Then, by equations \eqref{eq: frame compatibility}, we conclude that $R=0$, as required. Equations \eqref{eqs: lb compatibility}, \eqref{eqs: lpsi compatibility}, and \eqref{eqs: bpsi compatibility} can be interpreted as definitions of $\boldsymbol{K},\boldsymbol{\chi}$, and $\boldsymbol{\sigma}$, reducing the equations \eqref{eq: frame compatibility} to flatness conditions on the metric.

\section{Relation between $(l,\alpha)$ and Boozer coordinates} \label{apdx: Boozer coords}

In this appendix, we detail the transformation between the $(l,\alpha)$ coordinates used in this paper and Boozer coordinates $(\theta,\phi)$, where $\theta$ is the poloidal angle and $\phi$ is the toroidal angle. We do this for the particular case of QP symmetry with $2\pi I(\psi)=0$, where $I$ is the enclosed toroidal current for a flux surface with enclosed toroidal flux $2\pi\psi$. Due to $I=0$, $\boldsymbol{B}=G(\psi)\bnabla\phi + \beta(\psi,\theta,\phi) \bnabla \psi$ and the Jacobian in this coordinate system is $G(\psi)/B^2$. QP is then expressed as $B=B(\psi,\phi)$ (see, e.g., \citealt{helander_theory_2014}).

It can be shown that $\boldsymbol{u}=(\partial \boldsymbol{r}/\partial \theta)_{\phi,\psi}$ satisfies \eqref{eqs: governing eqs} for QP, where $B=B(\psi,\phi)$. It is then clear that $\boldsymbol{u}\cdot\boldsymbol{B}=0$. The arc length along the field line, with a particular choice of origin, is given by
\begin{equation}\label{eq: l to phi}
    l = \int_0^\phi d\phi' \frac{G(\psi)}{B(\psi,\phi')}.
\end{equation}
We also take $\alpha = \theta - \iota \phi$. Since $l=l(\psi,\phi)$, we then see that $\boldsymbol{u}=(\partial \boldsymbol{r}/\partial \alpha)_{l,\psi}$, as required. 

Also, as a corollary, we see that the length of a field line segment from $\phi=0$ to $\phi=2\pi$ is independent of the field line. This length is $L$, as defined in \S\ref{sec: possibility of irrational QP surfaces}.

Furthermore, we show why, given MHS force balance and $I=0$, the parallel current on the surface is zero. Under MHS force balance, $\beta$ has the same symmetry as the magnetic field strength (see, e.g., \citealt{Rodriguez2022}). So,
\begin{equation}
    \bnabla \times \boldsymbol{B} = \left(\frac{dG}{d\psi}  - \frac{\partial \beta}{\partial \phi} \right) \bnabla \psi \times \bnabla \phi
        \propto \frac{\partial \boldsymbol{r}}{\partial \theta}.
\end{equation}
Therefore, $\boldsymbol{J}$ is perpendicular to $\boldsymbol{B}$. This provides further support for why QP surfaces are natural extensions of LQP surfaces.

\section{Derivation of the surface equations}\label{apdx: derivation of surface equations}
In this appendix, we explicitly derive the surface equations \eqref{eqs: surface equations} from the relevant equations for a single LQP flux surface, as laid out in \S\ref{subsec: the surface equations}. We also derive equations \eqref{eqs: complete surface solution}, which are not overdetermined and accompany the surface equations in describing an LQP surface with all its relevant qualities.

To summarise the goal of this derivation, we desire $\lambda,\mu,\nu,\sigma,\boldsymbol{K}, \boldsymbol{\chi},\rho,\eta,\boldsymbol{K}_\psi, \boldsymbol{\chi}_\psi,\rho_\psi,$ and $\eta_\psi$ at fixed $\psi=\psi_0$ as functions of $(l,\alpha)$ that solve equations \eqref{eq: frame compatibility}, \eqref{eqs: lpsi compatibility}, \eqref{eqs: bpsi compatibility} at fixed $\psi=\psi_0$ and equations \eqref{eqs: lb compatibility} for $\psi=\psi_0+\delta \psi$ (the `geometric' constraints), as well as equations \eqref{eqs: physical constraints} (the `physical' constraints). Obtaining a general, analytical solution to these equations is not feasible. Instead, in this appendix, we algebraically reduce these equations to equations \eqref{eqs: surface equations} and \eqref{eqs: complete surface solution}. This reduction allows us to make claims about the surface equations being decoupled and complete, which are expressed in \S\ref{subsec: the surface equations}. In this appendix, variables are taken as functions of $(l,\alpha)$ at some fixed $\psi$ corresponding to the surface.

From equations \eqref{eq: lb compatibility a}, \eqref{eqs: physical constraints b}, and \eqref{eq: assumption parallel current weak}, we have
\begin{equation}\label{eq: A1}
    K_2 = 0, \quad K_{2\psi}=0,
\end{equation}
noting that equations \eqref{eqs: lb compatibility}
apply at $\psi_0$ and at $\psi_0 +\delta \psi $ for infinitesimal $\delta \psi$. Similarly, equation \eqref{eq: lb compatibility b} gives
\begin{equation}\label{eq: A2}
    \chi_3 = \rho \tau, \quad \chi_{3\psi}=(\rho \tau)_\psi,
\end{equation}
and equation \eqref{eq: lb compatibility c} gives
\begin{equation}\label{eq: A3}
    \chi_2 = \rho_l, \quad \chi_{2\psi}=\rho_{l \psi}.
\end{equation}
The second component of equation \eqref{eq: frame compatibility a} gives
\begin{equation}\label{eq: A4}
    \chi_1 = - \frac{\rho_{ll}-\rho \tau^2}{\kappa}.
\end{equation}
Though equations \eqref{eq: frame compatibility} hold at fixed $\psi=\psi_0$, the $\psi$ derivative of equation \eqref{eq: frame compatibility a} is satisfied automatically by equations \eqref{eq: frame compatibility b} and \eqref{eq: frame compatibility c}. Thus, we also have
\begin{equation}\label{eq: A5}
    \chi_{1\psi} = - \left( \frac{\rho_{ll}-\rho \tau^2}{\kappa} \right)_\psi.
\end{equation}
The first component of equation \eqref{eq: frame compatibility a} gives, after using the above expressions for $\boldsymbol{\chi}$, the second surface equation \eqref{eq: codazzi-mainardi b}, repeated here for convenience,
\begin{equation}\label{eq: A6}
    \tau_\alpha = \rho_l \kappa + \left( \frac{\rho_{ll}-\rho \tau^2}{\kappa} \right)_l,
\end{equation}
and similarly the third component of \eqref{eq: frame compatibility a} gives the first surface equation \eqref{eq: codazzi-mainardi a}, also repeated here,
\begin{equation}\label{eq: A7}
    \kappa_\alpha = -2 \rho_l \tau -\rho \tau_l.
\end{equation}
By the same reasoning that led to equation \eqref{eq: A5}, we also have the $\psi$ derivatives of equations \eqref{eq: A6} and \eqref{eq: A7}. But these equations are automatically satisfied due to equations \eqref{eq: frame compatibility}, so we do not write this explicitly. In other words, this result follows from our expressions for $\partial_\psi (\rho,\kappa,\tau)$ that we derive below.

Now, equation \eqref{eq: lpsi compatibility a} gives $\lambda_l = \mu \kappa$. We take the $\alpha$ derivative of this and use equations \eqref{eqs: physical constraints c} and \eqref{eq: assumption parallel current weak} to get the third surface equation,
\begin{equation}\label{eq: A8}
    \left( \frac{\kappa}{\rho} \right)_\alpha = 0.
\end{equation}

Directly from \eqref{eqs: physical constraints c}, we take $\mu$ as a solution to
\begin{equation}\label{eq: A9}
    \mu_\alpha = - \frac{\rho_\alpha}{\rho} \mu.
\end{equation}
Similarly, we take $\lambda$ as a solution to the compatible differential equations
\begin{equation}\label{eq: A10}
    \lambda_\alpha=0, \quad \lambda_l = \kappa \mu.
\end{equation}
 
We now solve equation \eqref{eq: lpsi compatibility b} for $\sigma_3$, which gives
\begin{equation}\label{eq: A11}
    \sigma_3 = \tau \nu - \kappa \lambda - \mu_l,
\end{equation}
and similarly, we solve equation \eqref{eq: lpsi compatibility c} for $\sigma_2$,
\begin{equation}\label{eq: A12}
    \sigma_2 = \nu_l + \tau \mu.
\end{equation}

Substituting this solution for $\sigma_2$ into equation \eqref{eqs: bpsi compatibility a} and using the fact that the parallel current vanishes (equation \eqref{eq: assumption parallel current weak}), we obtain
\begin{equation}\label{eq: A13}
    \nu_l = \frac{\rho_l}{\rho}\nu - 2 \tau \mu.
\end{equation}
We take $\nu$ to be a solution to this differential equation.

We still have to satisfy equations \eqref{eqs: bpsi compatibility b}, \eqref{eqs: bpsi compatibility c}, \eqref{eq: frame compatibility b}, and \eqref{eq: frame compatibility c}. 

We solve equation \eqref{eqs: bpsi compatibility b} for $\sigma_1$,
\begin{equation}\label{eq: A14}
    \sigma_1 = - \frac{\rho_\alpha}{\rho^2}\mu - \tau \lambda - \frac{\rho_{ll}-\rho \tau^2}{\kappa} \frac{\nu}{\rho},
\end{equation}
where we have substituted for $\sigma_3$ and used equation \eqref{eq: A9}. 

We solve equation \eqref{eqs: bpsi compatibility c} for $\rho_\psi$,
\begin{equation}\label{eq: A15}
    \rho_\psi = \nu_\alpha + \rho_l \lambda + \frac{\rho_{ll}-\rho \tau^2 }{\kappa}\mu,
\end{equation}
where we have substituted for $\boldsymbol{\chi}$ and $\boldsymbol{\sigma}$. 

We solve the first component of equation \eqref{eq: frame compatibility b} for $\tau_\psi$, which, after using the surface equations, reduces to
\begin{equation}\label{eq: A16}
    \tau_\psi = \frac{\tau_\alpha}{\rho}\nu + \tau_l \lambda + \frac{\rho_\alpha}{\rho^2}\mu_l + \left( \left(\frac{\rho_\alpha}{\rho^2} \right)_l - 2 \frac{\tau}{\rho} \frac{\rho_{ll}-\rho \tau^2}{\kappa}\right) \mu.
\end{equation}

Now, using the results we already have, we can in fact \textit{derive} the second component of equation \eqref{eq: frame compatibility b}. In particular, it follows from equations \eqref{eq: A1}, \eqref{eq: A11}, \eqref{eq: A13}, \eqref{eq: A14}, and the surface equations \eqref{eq: A6}, \eqref{eq: A7}, and \eqref{eq: A8}. Thus, equation \eqref{eq: frame compatibility b} gives no additional constraints.

We solve the third component of equation \eqref{eq: frame compatibility b} for $\kappa_\psi$, which gives, after using equations \eqref{eq: A10} and \eqref{eq: A13}, as well as the solutions for $\boldsymbol{\sigma}$,
\begin{equation}
    \kappa_\psi = \mu_{ll} + (\kappa^2 + 3 \tau^2) \mu + \kappa_l \lambda + \frac{\kappa_\alpha}{\rho}\nu. 
\end{equation}

Now we consider the first component of equation \eqref{eq: frame compatibility c}. For the first component of equation \eqref{eq: frame compatibility c}, we substitute the solutions for $\boldsymbol{\sigma}$ and $\boldsymbol{\chi}$, as well as for $\chi_{1 \psi}$ from equation \eqref{eq: A5}. In the substituted expression for $\chi_{1 \psi}$, we further substitute the solutions for $\rho_\psi$, $\kappa_\psi$, and $\tau_\psi$. In this way, it is verified (after much algebra) that the first component of equation \eqref{eq: frame compatibility c} is automatically satisfied; i.e., it can be derived from the results we already have and does not impose any further constraints. We can follow a similar process for the other two components of equation \eqref{eq: frame compatibility c} and, remarkably, we also find that these equations can be derived from the results we already have. 

Thus, we have successfully shown that the relevant equations and variables for defining an LQP flux surface (as laid out in \S\ref{sec: moving frame representation} and in equations \eqref{eqs: physical constraints}) reduce to the surface equations and the associated system of equations \eqref{eqs: complete surface solution}.

It is instructive to compare this calculation with that of \cite{SCHIEF_2003}. Our case turns out to be more tractable due to some fortunate cancellations. Namely, $\mu$ and $\nu$ are given as the solutions to individual differential equations (equations \eqref{eqs: complete surface solution b} and \eqref{eqs: complete surface solution c}) so there is no overdetermination in these variables and there are therefore no additional compatibility constraints. Remarkably, terms involving $\mu_l$, $\mu_{ll}$, and $\nu_\alpha$ cancel.

\section{Relationship to the results and language of classical differential geometry}\label{apdx: classical differential geometry}
In this appendix, we detail the relationship between the surface equations and some classical results of differential geometry. We note that there are no new results presented here: instead, this appendix seeks to express some typical concepts and results in terms of the surface quantities and equations.

We note that equations \eqref{eq: codazzi-mainardi a} and \eqref{eq: codazzi-mainardi b} came directly from the first two components of equation \eqref{eq: frame compatibility a}, after substituting for $\boldsymbol{\chi}$. Thus, they are purely geometric and satisfied by any embedded surface, written in geodesic coordinates. These two equations are known as the Codazzi-Mainardi equations \citep{Depman52,Mainardi1856,codazzi_sulle_1868}.

Given the surface quantities $( \kappa, \tau,\rho)$, we can, in principle, reconstruct the surface. This reconstruction is involved, as it requires integrating equations \eqref{eqs: defs for K, chi, sigma a} and \eqref{eqs: defs for K, chi, sigma b} and further integrating $\boldsymbol{r}_l = \boldsymbol{\hat{t}}$ and $\boldsymbol{r}_\alpha =\rho \boldsymbol{\hat{b}}$. 
Some geometric features of the surface are more easily calculated. The differential invariants of the surface are the Gaussian and mean curvature. By Gauss' (\citeyear{Gauss1828}) \textit{Theorema Egregium}, the Gaussian curvature is
\begin{equation}\label{eq: gaussian curvature}
    K = -\frac{\rho_{ll}}{\rho}.
\end{equation}
The mean curvature is
\begin{equation}
    H = \kappa - \frac{1}{\rho} \frac{\rho_{ll}-\rho \tau^2}{\kappa}.
\end{equation}

In the language of classical differential geometry (see, e.g., \citealt{eisenhart2013treatise}), the first fundamental form (also known as the metric or the line element) is
\begin{equation}\label{eq: surface metric in l alpha}
    \mathrm{I}=d\boldsymbol{r}\cdot d\boldsymbol{r}|_{\psi=\psi_0} = dl^2 + \rho^2 d\alpha^2.
\end{equation}
This determines the intrinsic geometry. The second fundamental form is
\begin{equation}
    \mathrm{I\!I}=-d\boldsymbol{r}\cdot d\boldsymbol{\hat{n}}|_{\psi=\psi_0} =\kappa \, dl^2 - 2\rho \tau \, dl \, d\alpha - \rho \frac{\rho_{ll} - \rho \tau^2}{\kappa} d\alpha^2  ,
\end{equation}
which controls the extrinsic geometry. By the Gauss-Bonnet theorem (see, e.g., \citealt{carmo_differential_1976}), the surface integral of the Gaussian curvature of a compact surface is equal to $2\pi\chi$ , where $\chi$ is the Euler characteristic of the surface. For the case of a toroidal surface, $\chi=0$, so
\begin{equation}\label{eq: Gauss-Bonnet for torus}
    0= \int  dA \, K = -\iint   dl d\alpha \, \rho_{ll}.
\end{equation}
For a QP surface, the Gauss-Bonnet theorem is an alternative expression of results that are already well known. The surface can be represented by a square grid in $(l,\alpha)$, since $\alpha$ can be scaled to be an angle-like quantity in the range $(0,2 \pi)$ and $l$ is in the range $(0,L)$ for $L$ independent of $\alpha$---this is a documented property of quasisymmetry (see, e.g., \citealt{helander_theory_2014} or \S\ref{sec: possibility of irrational QP surfaces}). Then, noting that $\rho$ is a single-valued quantity on the surface, the Gauss-Bonnet result \eqref{eq: Gauss-Bonnet for torus} is trivial. 

The quasisymmetry vector $\boldsymbol{u}=\partial_\alpha \boldsymbol{r}$ solves the Jacobi equation (see, e.g., \citealt{carmo_differential_1976})
\begin{equation}\label{eq: jacobi eqn}
    \frac{D^2 \boldsymbol{u}}{dl^2} + K \left( \boldsymbol{\hat{t}} \times \boldsymbol{u} \right) \times \boldsymbol{\hat{t}} = 0,
\end{equation}
where we have defined the covariant derivative
\begin{equation}
    \frac{D}{dl} = \boldsymbol{\hat{t}} \cdot \bnabla \boldsymbol{v}- \boldsymbol{\hat{n}}\boldsymbol{\hat{n}} \cdot \left( \boldsymbol{\hat{t}} \cdot \bnabla \boldsymbol{v} \right),
\end{equation}
where $\boldsymbol{v}$ is some vector in the surface's tangent plane. A foundational result of differential geometry is that the covariant derivative can be determined from the surface metric alone, through the Levi-Civita connection \citep{Christoffel+1869+46+70,levi-civita_nozione_1916}.

\section{Generalised helicoids as solutions to the surface equations}\label{apdx: generalized helicoids}

This appendix proves that generalised helicoids are the class of solutions to the surface equations when $\rho_\alpha=\kappa_\alpha=\tau_\alpha=0$. We also show how the generalised helicoid, along with surfaces of revolution and surfaces generated by a planar curve translated in the binormal direction, are consequently the only cases where $\boldsymbol{u}$ is a Killing vector of Euclidean space, i.e., the generator of helical, rotational, or translational motion. We note that generalised helicoids, except for the subcase of surfaces of revolution, have little practical application to the task of producing stellarator magnetic fields.

Generalised helicoids can be parameterised by
\begin{equation}\label{eq: generalized helicoid parameterization}
    \boldsymbol{r}(u,b) = \left(u \cos b, u \sin b, f(u) + a b \right),
\end{equation}
for arbitrary smooth function $f$ and constant $a$ \citep{eisenhart2013treatise}. This will be our starting point. 

Thus, we have
\begin{subequations}
    \begin{align}
        \frac{\partial \boldsymbol{r}}{\partial u} &= \left(\cos b, \sin b, f'(u) \right),\\
        \frac{\partial \boldsymbol{r}}{\partial b} &= \left( -u \sin b, u \cos b, a \right),
    \end{align}
\end{subequations}
which gives the surface metric, also known as the first fundamental form,
\begin{equation}
    \mathrm{I} = \left(1+f'(u)^2 \right) du^2 + 2af'(u) du db + \left(u^2 + a^2 \right)db^2.
\end{equation}
Compare this to the expression of the first fundamental form in $(l,\alpha)$ coordinates, equation \eqref{eq: surface metric in l alpha}. As well as being lines of inertial motion, geodesics are lines of shortest distance. So, the surface metric can be used to define an action from which we can apply the Lagrangian approach. In the Lagrangian sense, $b$ is an ignorable coordinate. Given that we parameterise the geodesic in terms of arc length, the geodesic is determined by the conservation laws,
\begin{subequations}\label{eqs: geodesic eqn for helicoid}
    \begin{align}
        1 &= \left(1+f'(u)^2 \right) \Dot{u}^2 + 2af'(u) \Dot{u} \Dot{b} + \left(u^2 + a^2 \right)\Dot{b}^2,\\
        J &= a f'(u) \Dot{u} + \left(u^2 + a^2 \right)\Dot{b},\label{eq: geodesic eqn for helicoid b}
    \end{align}
\end{subequations}
where $J$ is a constant of integration and the dot denotes a derivative with respect to $l$, the arc length along the geodesic.

The helices on the helicoid are given by $u=\text{const}$. In order for geodesics to be orthogonal to them, we take $J=0$. Then, using equation \eqref{eq: geodesic eqn for helicoid b}, we can define the coordinate
\begin{equation}
    \alpha = \int \! du \, \frac{a f'(u)}{u^2+a^2} + b,
\end{equation}
as is also defined in \cite{eisenhart2013treatise}. Thus, we have
\begin{equation}
    \mathrm{I} = \left( 1+ \frac{u^2 f'(u)^2}{u^2+a^2} \right)du^2 + (u^2 + a^2) d \alpha^2.
\end{equation}
We can then define the arc length as
\begin{equation}
    l = \int \! du \,\sqrt{1 + \frac{u^2 f'(u)^2}{u^2 + a^2}},
\end{equation}
to get
\begin{equation}
    \mathrm{I} = dl^2 + \rho^2 d\alpha^2,
\end{equation}
where $\rho=\sqrt{u^2 + a^2}$ is independent of $\alpha$, as required.

We now show that $\kappa$ and $\tau$ are independent of $\alpha$. There is a direct method: find $\Dot{u}$ and $\Dot{b}$ from equations \eqref{eqs: geodesic eqn for helicoid}, calculate $\boldsymbol{\hat{t}}=\Dot{u} (\partial \boldsymbol{r}/ \partial u)_b + \Dot{b}(\partial \boldsymbol{r}/\partial b)_u$, and calculate $\kappa$ and $\tau$ using the Frenet-Serret formulae. A quicker method uses the fact that the curvature $\kappa$ and torsion $\tau$ of a curve determine that curve uniquely, up to overall position and orientation (`the fundamental theorem of curves'). We therefore need only show that 
\begin{equation}\label{eq: condition for kappa tau independence}
    \left(\frac{\partial \boldsymbol{\hat{t}}}{\partial \alpha}\right)_l= \boldsymbol{v}(\alpha) \times \boldsymbol{\hat{t}},
\end{equation}
for some $\boldsymbol{v}(\alpha)$ that is independent of $l$. Using the geodesic equations \eqref{eqs: geodesic eqn for helicoid} we find
\begin{equation}
    \boldsymbol{\hat{t}} = \frac{1}{\sqrt{1+ \frac{u^2 f'(u)^2}{u^2+a^2}}}
        \begin{pmatrix}
            \cos b + \frac{a u f'(u)}{u^2+a^2} \sin b \\
            \sin b - \frac{a u f'(u)}{u^2+a^2} \cos b \\
            \frac{u^2 f'(u)}{u^2+a^2} \\
        \end{pmatrix}.
\end{equation}
Then, noting that $( \partial / \partial \alpha)_l = ( \partial / \partial b)_u$, we have equation \eqref{eq: condition for kappa tau independence} with $\boldsymbol{v}=(0,0,1)$. This completes the proof showing that $\kappa$ and $\tau$ are independent of $\alpha$.

We note that we have not explicitly shown that any solution where $\rho_\alpha=\kappa_\alpha=\tau_\alpha=0$ is a generalised helicoid. For this, we observe the fact that there is a helicoid corresponding to an arbitrary choice of $\rho,\kappa$, and $\tau$ as functions of $l$ (provided they obey the Codazzi-Mainardi equations) and that the Bonnet theorem applies. 

We now show why generalised helicoids, and their limiting cases, are the only solutions to the surface equations where $\boldsymbol{u}$ is a Killing vector of Euclidean space. To show this, we argue the stronger case that such surfaces are the only ones where $\boldsymbol{u}$, along a particular field line, can be extended to a Killing vector of Euclidean space. We reproduce a theorem from \cite{Langer1984}: a vector field $\boldsymbol{u}$ along a field line extends to a Killing vector in Euclidean space if and only if $\boldsymbol{u}$ satisfies
\begin{subequations}
    \begin{align}
        0 &= \boldsymbol{\hat{t}} \cdot \frac{d \boldsymbol{u}}{dl},\\
        0 &= \boldsymbol{\hat{n}} \cdot \frac{d^2 \boldsymbol{u}}{dl^2},\\
        0 &= \boldsymbol{\hat{b}} \cdot \left(\frac{d^3 \boldsymbol{u}}{dl^3} - \frac{\kappa_l}{\kappa} \frac{d^2 \boldsymbol{u}}{dl^2} + \kappa^2 \frac{d \boldsymbol{u}}{dl}\right),
    \end{align}
\end{subequations}
where $d/dl= \boldsymbol{\hat{t}}\cdot \bnabla$. By simple application of the Frenet-Serret formulae and the surface equations, these conditions are satisfied if and only if $\rho_\alpha=\kappa_\alpha=\tau_\alpha=0$. The Killing vectors of Euclidean space are either generators of translations, rotations, or helical motion. For translations, we get surfaces of the form $z=f(x)$, where $\boldsymbol{\hat{y}}$ is perpendicular to the field lines. For rotations, we get surfaces of revolution. For helical motion, we have generalised helicoids.

\section{QP objective functions for numerical optimisation}\label{apdx: qp objective targets}
For the first set of configurations (1--3), which were produced in {\tt DESC}, the QP objective function was
\begin{equation}\label{eq: DESC QP objective func}
    \frac{\int_0^{2 \pi} d\vartheta \int_0^{2 \pi} d\varphi \sqrt{g}_\text{surf}  \left| \boldsymbol{B}\times \bnabla B \cdot \bnabla \psi \right|}{\int_0^{2 \pi} d\vartheta \int_0^{2 \pi} d\varphi \sqrt{g}_\text{surf} B^3},
\end{equation}
where $\sqrt{g}_\text{surf}=\left| \partial_\theta \boldsymbol{r}\times \partial_\phi\boldsymbol{r}\right|$ is the Jacobian on the surface and $2\pi \psi$ is the toroidal magnetic flux. The expression \eqref{eq: DESC QP objective func} is the two-term error used in \cite{Dudt_Conlin_Panici_Kolemen_2023}.

For the second set of configurations (4--5), which were produced in {\tt SPEC} and optimised for using the adjoint methods in \cite{Nies_Paul_Hudson_Bhattacharjee_2022}, the QP objective function was
\begin{equation}
   \int_0^{2 \pi} d\vartheta \int_0^{2 \pi} d\varphi \sqrt{g}_\text{surf}  \left( \frac{1}{B^3}\boldsymbol{B}\times \bnabla B \cdot \bnabla \psi \right)^2.
\end{equation}
Because there are not necessarily nested flux surfaces within the boundary, defining $\bnabla \psi$ in this case is a subtle issue that is dealt with in \cite{Nies_Paul_Hudson_Bhattacharjee_2022}.

\section{Numerical calculation of $\rho$}\label{apdx: numerical calc of rho}
In this appendix, we show that the numerical calculation of $\rho$ as per equation \eqref{eq: numerical calculation of rho} agrees with the theoretical definition used in this paper.

The surface Jacobian $J$ in $(l,\alpha)$ coordinates is $J^{(l,\alpha)}=\rho$. Therefore, assuming a QP equilibrium, the surface Jacobian in Boozer coordinates is
\begin{equation}\label{eq: jacobian transform 1}
    J^{(\theta,\phi)}= \frac{G}{B}\rho,
\end{equation}
where we have used the transformation $(l=l(\theta,\phi),\alpha=\alpha(\theta,\phi))$ given in appendix \ref{apdx: Boozer coords}. Furthermore, using the standard transformation between Boozer coordinates and computational coordinates $(\vartheta,\varphi)$ given by equations \eqref{eqs: numerical transform to boozer}, we transform the surface Jacobian
\begin{equation}\label{eq: jacobian transform 2}
\sqrt{g}_\text{surf} \coloneq J^{(\vartheta,\varphi)}
    =\left|(1+\omega_\phi)(1+\tilde{\lambda}_\theta)-\omega_\theta (\tilde{\lambda}_\phi - \iota ) \right|J^{(\theta,\phi)}.
\end{equation}
Combining equations \eqref{eq: jacobian transform 1} and \eqref{eq: jacobian transform 2} gives equation \eqref{eq: numerical calculation of rho}.

\section{LQP surfaces that are perturbations to magnetic mirrors}\label{apdx: perturbations to surfs of rev}

\begin{figure}
    \centering
    \includegraphics[width=1.0\linewidth]{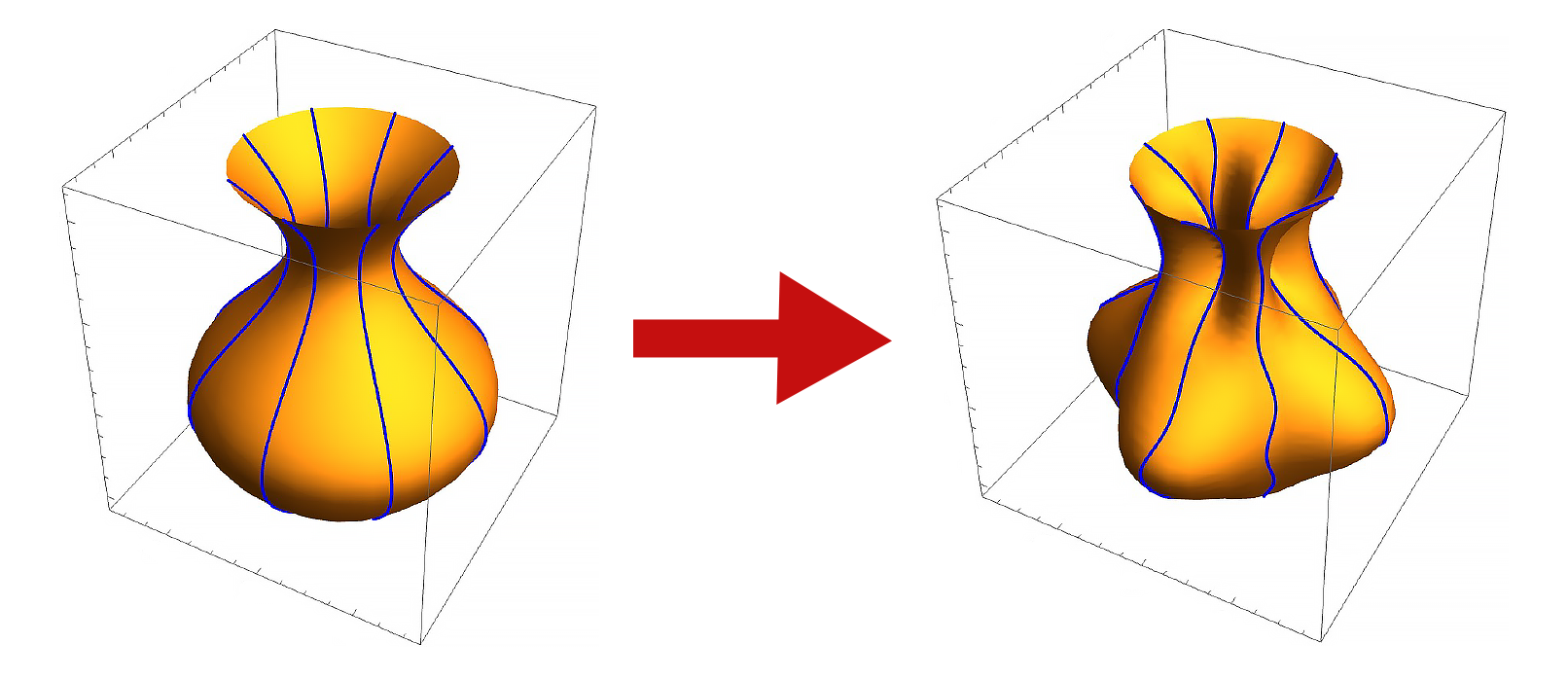}
    \caption{An illustrative picture of the perturbation to a magnetic mirror.}
    \label{fig:schematic perturbed mirror}
\end{figure}

In this appendix, we provide a mathematical argument based on the surface equations for why numerical QP equilibria tend to have narrow pinch points and high mirror ratios. This calculation is based on the identification of QP equilibria with `linked magnetic mirrors'. The calculation considers small perturbations to magnetic mirrors through the surface equations. We use this approach to derive the properties of the mirror-like surface that most readily permit the surface to `bend around' (as is necessary to form a toroidal surface) and generate a non-zero rotational transform. The schematic idea of perturbing a magnetic mirror is shown in figure \ref{fig:schematic perturbed mirror}. We now proceed through this calculation.

As the zeroth-order equilibrium surface solution, we take a surface of revolution with meridians as field lines:
\begin{equation}\label{eq: surf_of_rev parameterisation}
    \boldsymbol{r}(l,\alpha) = (x(l),y(l) \cos{\alpha}, y(l) \sin{\alpha}),
\end{equation}
where $l$ is the arc length of the field lines, so $x'^2+y'^2=1$. We use primes to signify a derivative with respect to $l$. For such a surface, the field lines have zero torsion (since they are planar), the (signed) curvature is $\kappa_0 = y''/x'$, and we also have $\rho_0=y$. We also define $\gamma \coloneq \kappa_0/\rho_0$. Defined in this way, the surface normal points inwards and $\alpha$ is an angle-like 
coordinate in the range $[0,2\pi)$. 

We now consider small perturbations to the equilibrium magnetic-mirror solution,
\begin{subequations}
    \begin{align}
        \rho &= y + \delta \rho,\\
        \kappa &= \kappa_0 + \delta \kappa,\\
        \tau &=  \delta \tau.
    \end{align}
\end{subequations}
Since we work with linear equations with coefficients that are independent of $\alpha$, it is useful to expand in Fourier modes of $\alpha$, 
\begin{equation}
    \delta \rho = \sum_{m=-\infty}^\infty \delta \rho_m (l) e^{i m \alpha},
\end{equation}
and similarly for $\delta \kappa$ and $\delta \tau$. We note that $m$ is restricted to integers, since $\alpha$ is an angle-like coordinate. For non-trivial results, we focus on the case where the perturbations are $\alpha$-dependent, so we ignore the $m=0$ mode. The $m=\pm1$ modes are particularly relevant for a toroidal geometry, since a positive $\delta \kappa$ on one side of the surface and a negative $\delta \kappa$ on the opposing side would lead to the surface `bending around'. This is akin to the Ackermann steering geometry for a turning car, where one wheel turns more than another. Additionally, non-zero $\delta \tau$ can be interpreted as the generator of a non-zero rotational transform $\iota$, since adding torsion is required for the meridians of a surface of revolution to curve around the surface.   

Linearising the third surface equation \eqref{eq: rho_div_kappa_constraint} leads to $\delta \kappa_m = \gamma \delta \rho_m$. Substituting for $\delta \kappa_m$ in the other surface equations gives
\begin{subequations}
    \begin{align}
        i m \gamma \delta\rho_m &= - \frac{\left( y^2 \delta \tau_m\right)'}{y},\\
        im \delta \tau_m &= \gamma \left( y \delta \rho\right)' + \left(\frac{1}{\gamma y} \left( \delta \rho_m'' - \frac{y''}{y} \delta \rho_m\right) \right)'.\label{eq: linearised surf eqns b}
    \end{align}
\end{subequations}
These two equations are equivalent to a fourth-order ODE in $l$, with $m$ as a parameter.

For tractability, we simplify this system by taking $\gamma \rightarrow \infty$ whilst holding $\gamma \delta \rho_m = \delta \kappa_m$ fixed. In this case, the second set of terms on the right-hand side of equation \eqref{eq: linearised surf eqns b} can be neglected, reducing the order of the differential equation system from four to two. This limit corresponds to a magnetic mirror with large axisymmetric curvature $\kappa_0\gg  d(\log y)/dl \sim d(\log\xi)/dl$. Then, the system of linearised equations is equivalent to
\begin{equation}\label{eq: sturm-liouville problem}
    \frac{1}{\gamma} \left(\gamma y^2 \xi_m' \right)' = m^2 \xi_m,
\end{equation}
where $\xi_m= y \delta \rho_m$ and $\delta \tau_m = - \frac{i}{m} \gamma \xi_m'$. Due to the fact that a stellarator surface closes up on itself, often with a further field-period symmetry, it is natural to consider a periodic surface of revolution and consider periodic perturbations. We take $L$ to be the distance along a field line across one period of the surface of revolution. With such boundary conditions, there exist two infinite, unbounded sequences of real eigenvalues $\{\lambda_{+n}\}$ and $\{\lambda_{-n}\}$ such that
\begin{subequations}
    \begin{align}
        0<\lambda_{+1}<\lambda_{+2}<\dots \rightarrow \infty,\\
        0>\lambda_{-1}>\lambda_{-2}>\dots \rightarrow -\infty,
    \end{align}
\end{subequations}
\citep{Haupt1915, Richardson18, Atkinson1987, Constantin1997}. We note that the negative sequence of eigenvalues is possible because equation \eqref{eq: sturm-liouville problem} is not a regular Sturm-Liouville problem, since $\gamma$ can switch signs and pass through zero. The square of an integer is, in general, not in the spectrum for any particular surface of revolution, recalling that the $\lambda = m^2 = 1$ case is the most relevant for toroidal geometries. Therefore, the surface needs to be finely tuned to admit perturbations.

Now, we are positioned to understand why narrow pinch points arise in the observed configurations. We note the result of \cite{Atkinson1987}, which states that the asymptotic form for $\lambda_{+n}$ as $n \rightarrow \infty$ is
\begin{equation}
    \lambda_{+n} \sim \frac{n^2 \pi^2}{\left( \int_0^L \frac{dl}{y}\right)^2}.
\end{equation}
Thus, the density of positive eigenvalues increases, in the asymptotic limit, if $y$ has a small minimum (i.e., the surface of revolution has a narrow pinch point). Notably, due to the term $1/y$ in the integrand of the denominator, a pinch point has a seemingly larger effect than scaling down the cross-sectional area of the magnetic mirror or increasing $L$. Thus, given a higher density of eigenvalues, we claim that smaller fine-tuned adjustments of the magnetic mirror are required to produce a perturbation with $\lambda=1$. Thus, \textit{it is `easier' for a magnetic mirror with a narrower pinch point to `bend around' whilst remaining an LQP surface}. Similarly, it is `easier' for a magnetic mirror with a narrower pinch point to have a non-zero rotational transform. In a toroidal equilibrium, narrow pinch points are also associated with larger magnetic field strengths, since field lines are compressed together. This leads to the observed high mirror ratios.

We caution that this argument is not rigorous as we employed several simplifying assumptions. Firstly, we neglected terms corresponding to finite $\kappa_0$ to reduce the order from four to two. Secondly, we assumed that the eigenvalues can be shifted slightly by small deformations of the magnetic mirror. This allowed us to conclude from an increased density of eigenvalues that it is `easier' for a magnetic mirror with a pinch point to have a $\lambda=1$ eigenvalue. Thirdly, we assumed that a higher density of eigenvalues in the asymptotic limit of $n\rightarrow \infty$ also relates to a higher density of eigenvalues around $\lambda = 1$.

\bibliographystyle{jpp}

\end{document}